\documentclass[a4paper,11pt]{article}
\pdfoutput=1 

\usepackage{jcappub}
\usepackage{graphicx}
\usepackage{dcolumn}
\usepackage{amssymb,amsmath,bm}
\usepackage{color}
\usepackage[dvipsnames,table]{xcolor}
\usepackage{xfrac}
\usepackage{aas_macros}
\usepackage{mathrsfs}
\usepackage{subcaption}
\usepackage{rotating}
\usepackage{chngcntr}
\usepackage{units}
\usepackage{multirow}
\usepackage{xspace}
\usepackage{xfrac}
\usepackage{mathtools}
\usepackage{comment}
\usepackage{soul}
\usepackage{gensymb}
\usepackage{multirow}

\newcommand{\bN}{{\bf N}}

\newcommand{\cN}{\mathcal{N}}
\newcommand{\Om}{\Omega_{\rm m}}

\newcommand{\ob}{\omega_{\rm b}}
\newcommand{\oc}{\omega_{\rm c}}

\newcommand{\As}{A_{\rm s}}
\newcommand{\ns}{n_{\rm s}}

\newcommand{\enangle}[1]{\langle #1 \rangle}

\newcommand{\md}{\mathrm{d}}

\newcommand{\mlim}{m_{\rm lim}}
\newcommand{\sky}{{\rm sky}}

\newcommand{\fsky}{f_{\rm sky}}

\newcommand{\shear}{\boldsymbol{\gamma}}

\newcommand{\cov}{{\sf Cov}}

\newcommand{\zbins}{$z$-bins\xspace}

\newcommand{\nmt}{\texttt{NaMaster}\xspace}
\newcommand{\cobaya}{\texttt{Cobaya}\xspace}
\newcommand{\healpix}{\texttt{HEALPix}\xspace}
\newcommand{\ccl}{\texttt{CCL}\xspace}

\newcommand{\camb}{\texttt{CAMB}\xspace}
\newcommand{\hfit}{\textsc{HALOFIT}\xspace}

\newcommand{\des}{DES\xspace}
\newcommand{\desyo}{DESY1\xspace}
\newcommand{\hsc}{HSC\xspace}
\newcommand{\hscdro}{HSC-DR1\xspace}
\newcommand{\mcal}{\textsc{Metacalibration}\xspace}
\newcommand{\planck}{{\sl Planck}\xspace}
\newcommand{\cosmos}{COSMOS2015\xspace}

\usepackage{natbib}
\title{Combining cosmic shear data with correlated photo-$z$ uncertainties: constraints from \desyo and \hscdro}

\author[a,1]{Carlos Garc\'ia-Garc\'ia,}
\author[a]{David Alonso,}
\author[a]{Pedro G. Ferreira,}
\author[b,c,d]{Boryana Hadzhiyska,}
\author[e,f]{Andrina Nicola,}
\author[g,h]{Carles S\'anchez,}
\author[i]{An\v ze Slosar}

\affiliation[a]{Department of Physics, University of Oxford, Denys Wilkinson Building, Keble Road, Oxford OX1 3RH, United Kingdom}
\affiliation[b]{Harvard-Smithsonian Center for Astrophysics, 60 Garden St, Cambridge, MA 02138, USA}
\affiliation[c]{Miller Institute for Basic Research in Science, University of California, Berkeley, CA, 94720, USA}
\affiliation[d]{Physics Division, Lawrence Berkeley National Laboratory, Berkeley, CA 94720}
\affiliation[e]{Department of Astrophysical Sciences, Princeton University, Peyton Hall, Princeton, NJ 08544, USA}
\affiliation[f]{Washington University in St. Louis, Physics Department, 1 Brookings Drive, St. Louis, MO 63130, USA}
\affiliation[g]{Institute of Space Sciences (ICE, CSIC), Campus UAB, Carrer de Can Magrans, 08193 Barcelona, Spain}
\affiliation[h]{Department of Physics \& Astronomy, University of Pennsylvania, 209 S. 33rd St., Philadelphia, PA 19104, USA}
\affiliation[i]{Physics Department, Brookhaven National Laboratory, Upton NY 11973, USA}

\emailAdd{carlos.garcia-garcia@physics.ox.ac.uk}

\abstract{An accurate calibration of the source redshift distribution $p(z)$ is a key aspect in the analysis of cosmic shear data. This, one way or another, requires the use of spectroscopic or high-quality photometric samples. However, the difficulty to obtain colour-complete spectroscopic samples matching the depth of weak lensing catalogs means that the analyses of different cosmic shear datasets often use the same samples for redshift calibration. This introduces a source of statistical and systematic uncertainty that is highly correlated across different weak lensing datasets, and which must be accurately characterised and propagated in order to obtain robust cosmological constraints from their combination. In this paper we introduce a method to quantify and propagate the uncertainties on the source redshift distribution in two different surveys sharing the same calibrating sample. The method is based on an approximate analytical marginalisation of the $p(z)$ statistical uncertainties and the correlated marginalisation of residual systematics. We apply this method to the combined analysis of cosmic shear data from the \desyo data release and the \hscdro data, using the COSMOS 30-band catalog as a common redshift calibration sample. We find that, although there is significant correlation in the uncertainties on the redshift distributions of both samples, this does not change the final constraints on cosmological parameters significantly.  The same is true also for the impact of residual systematic uncertainties from the errors in the COSMOS 30-band photometric redshifts. Additionally, we show that these effects will still be negligible in Stage-IV datasets. Finally, the combination of \desyo and \hscdro allows us to constrain the ``clumpiness'' parameter to $S_8 = 0.768^{+0.021}_{-0.017}$. This corresponds to a $\sim\sqrt{2}$ improvement in uncertainties with respect to either DES or HSC alone.}

\begin{document}
\maketitle

\section{Introduction}
  The gravitational lensing of background galaxies is one of the most reliable probes to study the late-time growth of structure, as well as the geometry of our Universe. Imaging weak lensing surveys such as the Dark Energy Survey (DES) \citep{2105.13543,2203.07128}, the Hyper Suprime-Cam (HSC) \citep{1809.09148,1906.06041}, and the Kilo-Degree Survey (KiDS) \citep{2007.15632}, have produced increasingly tight direct constraints on the amplitude of matter density inhomogeneities, and on the abundance of non-relativistic matter at $z\lesssim1$. Although qualitatively in line with the standard cosmological model ($\Lambda$CDM) favoured by observations of the CMB at early times, these measurements are quantitatively in tension with constraints from \planck \cite{1502.01589, 1807.06209} at the $\sim3\sigma$ level \cite{1906.09262, 2007.15633, 2105.12108}. If real, this tension could be a smoking gun signature of new physics \cite{1711.00692, 2105.09249, 2207.03500., 2209.06217, 2209.08102}, although it might also be explained by a poor understanding of baryonic effects \cite{2202.07440, 2206.11794}, systematic uncertainties in the characterisation of the source redshift distribution \cite{1906.09262}, or unknown systematic effect in the CMB measurements. These experiments will be superseded within the next decade by so-called ``Stage-IV'' surveys, such as the ground-based Rubin Observatory Legacy Survey of Space and Time (LSST) \citep{0805.2366,1809.01669}, the Euclid satellite mission \citep{2012SPIE.8442E..0TL}, and the Nancy Grace Roman Space Telescope \citep{1902.05569}. These surveys will improve the width, depth, and angular resolution of current experiments, and will be vital in ascertaining the source of this tension.

  Carrying out combined analyses that add the statistical power of different experiments \citep{1808.07335,1906.09262,2105.12108,2111.09898,2202.07440,2204.10392,2208.07179} is key to improving their individual cosmological constraints, and to test the robustness of these constraints to instrumental or observational systematics. This is often challenging when surveys targeting potentially different source populations share part of the sky, due to the non-trivial survey geometry effects, and the need to characterise the noise properties of the sources common to both datasets. However, even disjoint observations may share correlated systematic and statistical uncertainties due to the use of common external datasets, which must be carefully modelled and propagated. This is the case of photometric redshift uncertainties.

  A precise calibration of the redshift distribution $p(z)$ of the source sample is a key component of the cosmological analysis. To achieve it, weak lensing surveys rely on the use of external spectroscopic or high-quality photometric datasets. The redshift distribution can then be estimated via direct calibration or through a clustering redshifts approach. Direct calibration involves using directly the redshift distribution of the spectroscopic sample. Since the spectroscopic sample is often not representative of the sample used to measure the weak lensing signal, care must be taken to re-weight the former to correct for its different color-magnitude distribution. A variety of methods have been proposed to do so \cite{0801.3822, 1909.09632}. In the clustering redshifts method, the weak lensing sample is cross-correlated with the spectroscopic sample in narrow bins of redshift. The amplitude of the resulting cross-correlation is proportional to the redshift distribution of the photometric sample. Care must be taken then to account for the evolution of the photometric sample's bias within the redshift range of interest, and to quantify the potential contamination from magnification bias in the cross-correlation \cite{astro-ph/0606098,0805.1409,1003.0687,1303.0292,2012.08569}. Another promising method that can help calibrate weak lensing redshift distributions is that of shear-ratios~\cite{1708.01537,2105.13542}, especially when used in combination with 3x2pt measurements. In this case, one measures the spacial averaged ratio of the galaxy-galaxy lensing between the same galaxy sample and two galaxy lensing \zbins, which mainly depends on the redshift distributions of the three samples and the cosmological parameters. However, regardless of how the $p(z)$ is measured, since large spectroscopic samples of the required depth are often difficult to come by, it is likely that different weak lensing surveys will make use of the same (or at least partially overlapping) spectroscopic samples. Thus, the statistical uncertainties in the determination of redshift distributions, as well as any systematic trends in the shared spectroscopic sample, will be correlated between different cosmic shear analyses. As an example of this, the analysis of the first-year data from the Dark Energy Survey (\desyo hereafter), and the first data release from HSC (\hscdro hereafter), both relied heavily on data from the COSMOS 30-band photometric catalog of \citep{1604.02350} (\cosmos hereafter) for $p(z)$ calibration. Likewise, it is reasonable to expect that future surveys will aim to use the best spectroscopic dataset available at the time of their analysis (e.g. from the Dark Energy Spectroscopic Instrument \cite{1907.10688}), and thus the correlation in the $p(z)$ uncertainties of different datasets will need to be carefully quantified when comparing or combining their results. This will be important even in the (probably more common) case of having different overlapping areas. In such a case, a remaining systematic (correlated) error coming from the training set could correlate the calibration. This is something we investigate in this work in the context of the high-quality photometric datasets of \cosmos and PAUS~\cite{2007.11132}.

  In this paper, we make use of the analytical marginalisation scheme proposed in \citep{2007.14989} to address this problem, developing a method to consistently propagate $p(z)$ uncertainties for cosmic shear datasets that are otherwise independent. We apply this method to public data from DESY1 and HSC DR1 in order to combine the analysis of both datasets, quantifying the impact of correlated redshift calibration uncertainties.

  The paper is structured as follows. Section \ref{sec:meth} presents the cosmological theory background underlying the analysis of cosmic shear surveys, as well as the method used here to calibrate redshift distributions and propagate their uncertainties. Section \ref{sec:data} briefly describes the datasets used in our analysis. The main results from the analysis, both in terms of the impact of correlated $p(z)$ uncertainties, and the final combined cosmological constraints from DESY1 and HSC DR1 are presented in Section \ref{sec:results}, whereas in Section~\ref{sec:forecast} we study the expected effect in Stage-IV surveys. Finally, we conclude in Section \ref{sec:conc}.

\section{Methods}\label{sec:meth}
  \subsection{Theory}\label{ssec:meth.theory}
    The perturbation to observed ellipticities of a sample of galaxies with redshift distribution $p_\alpha(z)$ due to weak gravitational lensing is captured by the spin-2 ``cosmic shear'' field $\shear^\alpha_G$. The $E$-mode component of this field is then related to the three-dimensional matter overdensity via a line-of-sight projection with a radial kernel given by \cite{astro-ph/9912508}
    \begin{equation}
      q^\alpha_{\shear_G}(\ell, \chi) = \frac{3}{2} H_0^2 \Om \frac{\chi}{a(\chi)} \int_{z(\chi)}^\infty \md z' p_\alpha(z') \frac{\chi(z') - \chi}{\chi(z')},
    \end{equation}
    where $H_0$ the expansion rate, and $\Om$ is the fractional energy density of non-relativistic matter, both evaluated today. $z(\chi)$ is the redshift corresponding to a source at comoving radial distance $\chi$.
    
    The angular power spectrum between the cosmic shear signal from two different samples, $\alpha$ and $\beta$, is then related to the matter power spectrum $P(k,z)$ via
    \begin{equation}
      C_\ell^{\alpha\beta}=G_\ell^2\int\frac{d\chi}{\chi^2}q_{\shear_G}^\alpha(\chi)q_{\shear_G}^\beta(\chi)\,P\left(k=\frac{\ell+1/2}{\chi},z(\chi)\right),
    \end{equation}
    where we have made use of the Limber approximation \cite{1953ApJ...117..134L}, which holds for broad radial kernels (as is the case for cosmic shear). The scale-dependent prefactor
    \begin{equation}
        G_\ell\equiv\sqrt{\frac{(\ell+2)!}{(\ell-2)!}}\frac{1}{(\ell+1/2)^2}
    \end{equation}
    accounts for the relation between the cosmic shear field and the angular Hessian (as opposed to the three-dimensional Laplacian) of the Newtonian gravitational potential \cite{1702.05301}.

    The intrinsic correlation between true galaxy shapes, caused by local interactions, is a potential contaminant to the cosmic shear signal from gravitational lensing. Here we will model this ``intrinsic alignment'' (IA) contribution, $\shear_I$, using the ``linear-non-linear alignment model'' of \citep{astro-ph/0406275, 0705.0166},  where the perturbed galaxy ellipticity is proportional to the local gravitational tidal forces. This amounts to including an additive contribution to the radial kernel of the form:
    \begin{equation}
      q_{\shear_I}^\alpha(\chi)=-A_{\rm IA}(z)p_\alpha(z)\frac{dz}{d\chi}.
    \end{equation}
    We parametrise the IA amplitude as in \citep{1708.01530,1708.01538}
    \begin{equation}\label{eq:ia}
      A_{\rm IA} = A_{{\rm IA}, 0} \left(\frac{1+z}{1+z_0}\right)^{\eta_{\rm IA}} \frac{0.0139\Om}{D(z)}.
    \end{equation}
    Here, $D(z)$ is the linear growth factor normalised to $D(0)=1$, $z_0$ is a pivot redshift that we fix to $z_0=0.62$ as in \citep{1708.01530,1708.01538}, and $A_{{\rm IA}, 0}$ and $\eta_{\rm IA}$ are free parameters.

    We use the Core Cosmology Library (\ccl) \cite{1812.05995} to compute all angular power spectra, and \camb{} \cite{astro-ph/9911177} with \hfit \cite{astro-ph/0207664} to compute the non-linear matter power spectrum.

  \subsection{Measuring the $p(z)$}\label{ssec:meth.nz_cal}
    A calibrated model for the source redshift distribution $p(z)$ is a central component of the theoretical prediction for the measured shear power spectra. Here we use the direct calibration (DIR) method of \citep{0801.3822}. Consider two galaxy samples: a {\sl photometric} sample $P$ containing the photometry ${\bf m}$ (magnitudes measured in a set of $N_b$ bands) of each source, and a {\sl spectroscopic} sample $S$, which contains accurate redshift measurements in addition to ${\bf m}$ for each source. We will assume that $P$ is representative of the sample of galaxies used in the cosmic shear analysis (i.e. both follow the same distribution of ${\bf m}$). In practice, $S$ may be initially an external catalog that does not contain the same photometry as the photometric sample. In that case, photometry information is obtained by spatially cross-matching $S$ with another sample (not necessarily the same as $P$) containing the desired magnitude measurements. The DIR method then proceeds as follows:
    \begin{enumerate}
      \item For each object $i$ in $S$, we find its $N_{S}$ nearest neighbors in $S$, in the $N_b$-dimensional space of ${\bf m}$ using a Euclidean metric to determine distances. We then record the distance $d$ to the furthest neighbor. Our fiducial results use $N_{S}=20$ nearest neighbors but we have made sure they are not affected by this choice.
      \item For the same object in $S$, we find the number of objects $N_{P,i}$ in $P$ that lie within an $N_b$-sphere of the same radius $d$. If any weights were applied to the cosmic shear catalog in the cosmological analysis (e.g. shape measurement weights), $N_{P,i}$ should be the sum of weights. 
      \item We then assign a weight $w_i=r\,N_{P,i}/N_{S}$ to each object in $S$, where $r$ is the ratio between the total number of objects in $S$ and  the total (weighted) number of objects in $P$. $r$ is simply a normalisation constant that does not affect the final inferred redshift distribution.
      \item Finally, the redshift distribution of $P$ is estimated as the redshift distribution of $S$ where each source is weighted by $w_i$ defined above.
    \end{enumerate}
    
    It is worth noting that the DIR method is only guaranteed to provide an unbiased estimate of the redshift distribution under somewhat idealised conditions. First, although $S$ need not follow exactly the same color distribution as the target sample $P$, it must contain a sufficiently high number of objects in all regions of magnitude space where $P$ has support. Secondly, the redshift measurements in $S$ are assumed to be precise and accurate. In particular, for fixed photometry, the redshift success rate should not depend on the redhift itself. This is rarely true in practice, particularly at high redshifts and for faint sources. For example, the \cosmos catalog we will use in this work, uses photometric redshift measurements estimated from 30-band photometry. Although the resulting redshifts are significantly more accurate than photo-$z$s estimated from $O(4-5)$ bands, we can expect a number of catastrophic outliers, particularly at high redshifts \cite{1906.09262}. Finally, if $S$ is limited to a small sky patch (as is also the case for \cosmos), the $p(z)$ estimated via DIR will be subject to significant sample variance uncertainties that may be difficult to model in practice. In summary, even after estimating the source redshift distributions and their statistical uncertainties, we must allow for potential residual systematic deviations in the $p(z)$.

  \subsection{Marginalisation over $p(z)$ statistical uncertainties}\label{ssec:meth.nz_stat}
    We marginalize analytically over the uncertainty on the $p(z)$ following \cite{2007.14989}. The method is based on linearising the dependence of the model on the deviations with respect to a fiducial $p(z)$, which then allows us to marginalise over these deviations analytically. This marginalisation results in an additive contribution to the covariance matrix of the data vector (i.e. the set of angular power spectra), which effectively increases the uncertainty of the modes that are most affected by $p(z)$ fluctuations.
    
    The method proceeds as follows. Let ${\bf N}$ be a vector of parameters consisting of the amplitudes of the redshift distributions for all the galaxy samples analysed (e.g. the histogram heights for each $p(z)$ estimated via DIR). Let $\bar{\bf N}$ be the best-fit measurements of ${\bf N}$, and ${\sf P}$ the covariance matrix of these measurements. ${\sf P}$ may incorporate both statistical uncertainties and any prior constraints (e.g. imposing smoothness of the underlying redshift distribution). We will only consider statistical uncertainties here. Finally, let ${\sf T}$ be the matrix characterising the response of the data vector to variations in ${\bf N}$:
    \begin{equation}
      {\sf T}\equiv\frac{d{\bf C}}{d{\bf N}},
    \end{equation}
    where ${\bf C}$ is the vector containing all angular power spectra being analysed. Then, analytically marginalising over ${\bf N}$, assuming a linear dependence on the deviations with respect to $\bar{\bf N}$, is equivalent to modifying the covariance matrix of ${\bf C}$ as:
    \begin{equation}
      \cov_{\bf C}\rightarrow\cov_{\bf C}+{\sf T}{\sf P}{\sf T}^T.
      \label{eq:cov}
    \end{equation}
    
    As in ~\cite{2004.09542}, we use an analytical estimate of the covariance of ${\bf N}$. The matrix ${\sf P}$ has two contributions: the covariance $\cov^\delta$ due to fluctuations in the matter density traced by the galaxies in the observed patch and a shot-noise contribution $\cov^S$; i.e.
    \begin{equation}\label{eq:P}
        {\sf P} = \cov^\delta + \cov^S.
    \end{equation}
    The shot noise part is simply given by
    \begin{equation}
        \cov^S_{ij} = \delta_{ij} N_i,
    \end{equation}
    where $N_i$ is the number of sources in the corresponding histogram bin. The sample variance contribution is
    \begin{equation}
        \cov^\delta_{ij} = \frac{N_i N_j}{2\pi^2} \int_0^\infty dk_\parallel \cos(k_\parallel(\chi_i-\chi_j))\int_0^\infty dk_\perp k_\perp W_i(k_\parallel,k_\perp)W_j(k_\parallel,k_\perp)P_{gg}({\bf k})\,.
    \end{equation} 
    In this expression, $k_\parallel$ and $k_\perp$ are the components of the Fourier-space wavevector ${\bf k}$ parallel and perpendicular to the line of sight, and $\chi_i$ is the radial comoving distance to redshift $z_i$. $W_i$ is the Fourier-space window function for the 3D volume covered by each redshift interval over which ${\bf N}$ has been measured. Finally, $P_{gg}$ is the galaxy power spectrum in redshift space, which we model as \citep{1987MNRAS.227....1K}
    \begin{equation}
        P_{gg}({\bf k},z)=\left(b_g(z)+f(z)\frac{k_\parallel^2}{k^2}\right)^2\,P_{mm}(k,z).
    \end{equation}
    Here, $f(z)$  is the linear growth rate and $b_g(z)$ is the linear galaxy bias. For $b_g$, we use the fit of \cite{1912.08209} for flux-limited samples, which takes the form
    \begin{equation}
      b_g(z) = (b_1(\mlim - 24) + b_0) D(z)^\alpha,
      \label{eq:bg}
    \end{equation}
    with $b_1 = -0.0624$, $b_0 = 0.8346$, and $\alpha = -1.30$. $\mlim$ is the limiting magnitude in the $i$ band. We use $\mlim=24.5$ for HSC and $\mlim=22.5$ for DES. This fitting function was derived from the HSC galaxy clustering data, but we have verified that using this fit or fitting the bias directly to the galaxy clustering cross-correlation of the DES sample with Planck 18 CMB lensing~\cite{1807.06210}\footnote{The galaxy clustering fields from the DES weak lensing samples are computed considering the official weak lensing redshift distributions and the number counts and overdensity field as given by the galaxy position in the \mcal catalog for the galaxies lying in the corresponding redshift bin. Finally, the mask used is that of the redMaGiC sample~\cite{1708.01536} and the covariance is approximated by the Knox formula~\cite{astro-ph/9504054}.} leads to negligible differences in our final analysis once the diagonal of $P$ is enlarged to match the official errors.

    We model the window function $W_i$ as that of a top-hat cylinder with a comoving width $\Delta\chi$ given by the redshift interval used to define the redshift distribution histogram, and a radius $R=\chi\theta_{\rm sky}$, where $A_{\rm sky}=\pi\theta_{\rm sky}^2=2\,{\rm deg}^2$ is the area of the COSMOS field used in our analysis. The resulting functional form for $W_i$ is
    \begin{equation}
      W_i(k_\parallel,k_\perp)=j_0(k_\parallel\Delta\chi_i/2)\,\frac{2 J_1(k_\perp \chi_i\theta_{\rm sky})}{k_\perp \chi_i\theta_{\rm sky}},
    \end{equation}
    where $j_0$ is the zeroth-order spherical Bessel function and $J_1$ is the first-order cylindrical Bessel function. Since $\theta_{\sky} \propto \sqrt{A_{\sky}}$, a small change on $A_{\rm sky}$ barely changes $\theta_{\rm sky}$ and, therefore, $W_i$.
    
    Finally, in order to avoid underestimating the errors due to the analytical treatment used here, we allow ourselves the possibility to enlarge the diagonal of ${\sf P}$ by a factor $f_P$. The role of $f_P$ will be discussed in Section \ref{ssec:results.pz}.

  \subsection{Marginalisation over $p(z)$ systematic uncertainties}\label{ssec:meth.nz_syst}
    As we argued earlier, the fiducial redshift distribution $\bar{\bf N}$ may be subject to systematic uncertainties due to imperfections in the calibration sample. A particular concern in our case is the use of photometric redshifts in the \cosmos catalog. Refs. \citep{1906.09262, 2003.10454} have argued that this could cause a coherent shift in the means of all redshift distributions for a given survey. If a model for these residual systematics can be obtained (e.g. through the use of clustering redshifts, or from a higher-quality spectroscopic sample), their impact can be taken into account by marginalising over the free parameters of this model. A specific example of this will be described in Section \ref{ssec:results.pz_syst}, where we will use the COSMOS-PAUS catalog of \cite{2007.11132} to characterise a coherent redshift shift in the \cosmos catalog.

    When combining two different surveys calibrated with the same sample, these systematics will also be correlated. The effect, however, may not be the same for both surveys, since their source samples will in general cover different regions of color space. We can, however, explore the two limiting regimes of perfect correlation and complete independence. In the first case, we use the same nuisance parameters to characterise the systematic shift in the redshift distributions of both samples, whereas in the latter we use an independent set of parameters for each survey. If the final cosmological constraints are sensitive to this choice, a more sophisticated model for the correlation between the systematic effect in both surveys is necessary. In practice, we will show that this is not the case.

  \subsection{Likelihood}\label{ssec:meth.like}
    To extract parameter constraints from measurements of the cosmic shear power spectra in different surveys, we make use of a Gaussian likelihood. This choice is justified by the central limit theorem on the scales used in this analysis \cite{0801.0554}. The power spectrum measurements and corresponding covariance are described in Section \ref{ssec:data.cls}.

    Our model will depend on the 5 flat $\Lambda$CDM parameters $\omega_c=\Omega_ch^2$, $\omega_b=\Omega_bh^2$, $h$, $\log(A_s)$ and $n_s$. Here $\Omega_c$ and $\Omega_b$ are the fractional abundances of cold dark matter and baryons respectively, $h\equiv H_0/(100\,{\rm km}\,{\rm s}^{-1}\,{\rm Mpc}^{-1})$, $A_s$  is the primordial amplitude of scalar perturbations, and $n_s$ is the primordial scalar spectral index. The optical depth to reionisation was set to $\tau=0.08$, and we only considered massless neutrinos. Beside these cosmological parameters, we marginalised over 4 intrinsic alignment parameters, $\{A_{{\rm IA},0}^{\rm DES},A_{{\rm IA},0}^{\rm HSC},\eta_{\rm IA}^{\rm DES},\eta_{\rm IA}^{\rm HSC}\}$, characterising the effect of IAs in the DES and HSC samples according to Eq. \ref{eq:ia}. We also marginalise over a multiplicative bias parameter for each redshift bin, using the shape measurement priors quoted in \citep{1810.02322} and \citep{1809.09148} for \desyo and \hscdro respectively. When marginalising over residual systematic uncertainties in the calibrated redshift distributions, we will make use of a two-parameter linear model. The details, including the procedure used to obtain a prior on the parameters of the model, is described in Section \ref{ssec:results.pz_syst}. Finally, when comparing with the official \desyo and \hscdro constraints, we will marginalise over shifts in the mean of the redshift distribution of each redshift bin. The priors used on these nuisance parameters are listed in Table \ref{tab:priors}. For the cosmological parameter priors, we follow the \hscdro analysis \citep{1809.09148}. We found that this results in more conservative constraints (i.e. broader posteriors) than using the \desyo priors \citep{1708.01538}.
    
    We use \cobaya~\cite{2005.05290,2019ascl.soft10019T} to sample the parameter space with a Monte-Carlo Markov-Chain (MCMC) method. In particular, we use the Metropolis-Hasting~\cite{metropolismc,hastingsmc} algorithm and check convergence requiring the Gelman-Rubin~\cite{gelmanrubin} parameter to satisfy $R-1 < 0.01$ in the diagonalised parameter space.

    \begin{table}
        \centering
        \def\arraystretch{1.2}
        \begin{tabular}{|lr|}
        \hline
        Parameter &  Prior\\  
        \hline 
        $\oc$  &  $U (0.03, 0.7)$                             \\ 
        $\ob$  &  $U (0.019, 0.026)$                           \\
        $h$   &  $U (0.6, 0.9)$                             \\
        $\log(10^{10} \As)$ &  $U (1.5, 6.0)$             \\
        $\ns$ & $U (0.87, 1.07)$                             \\
        $\tau$  & 0.08                                       \\
        $\sum m_\nu$ & 0                                     \\ 
        \hline
        \end{tabular}
        \begin{tabular}{|l|cc|}
            \hline
                        & DES & HSC \\
            \hline
            $m_i$         & $\cN(0.012, 0.023)$ & $\cN(0, 01)$\\
            \hline
            $A_{{\rm IA}, 0}$ & $U(-5, 5)$ & $U(-5, 5)$\\
            $\eta_{\rm IA}$ & $U(-5, 5)$ & $U(-5, 5)$\\
            \hline
            $\Delta z_0$ &  $\cN(0., 0.016)$ & $\cN(0., 0.037)$\\
            $\Delta z_1$ &  $\cN(0., 0.017)$ & $\cN(0., 0.017)$\\
            $\Delta z_3$ &  $\cN(0., 0.013)$ & $\cN(0., 0.021)$\\
            $\Delta z_4$ &  $\cN(0., 0.015)$ & $\cN(0., 0.017)$\\
            \hline
        \end{tabular}
        \caption{Prior distributions for the cosmological (left) and nuisance (right) parameters. $U(a, b)$ and $\cN(\mu, \sigma)$ are a uniform distribution with boundaries $(a, b)$, and a Gaussian distribution with mean $\mu$ and variance $\sigma$, respectively. The $\Delta z_i$ have been calibrated for our $p(z)$ sampling from a multivariate Gaussian with mean and covariance $\bN$ and $P$ (see Eq.~\ref{eq:P}).} 
        \label{tab:priors}
    \end{table}

\section{Data}\label{sec:data}
  \subsection{DES and HSC}\label{ssec:data.deshsc}
\begin{figure}
      \centering
      \includegraphics[width=0.8\textwidth]{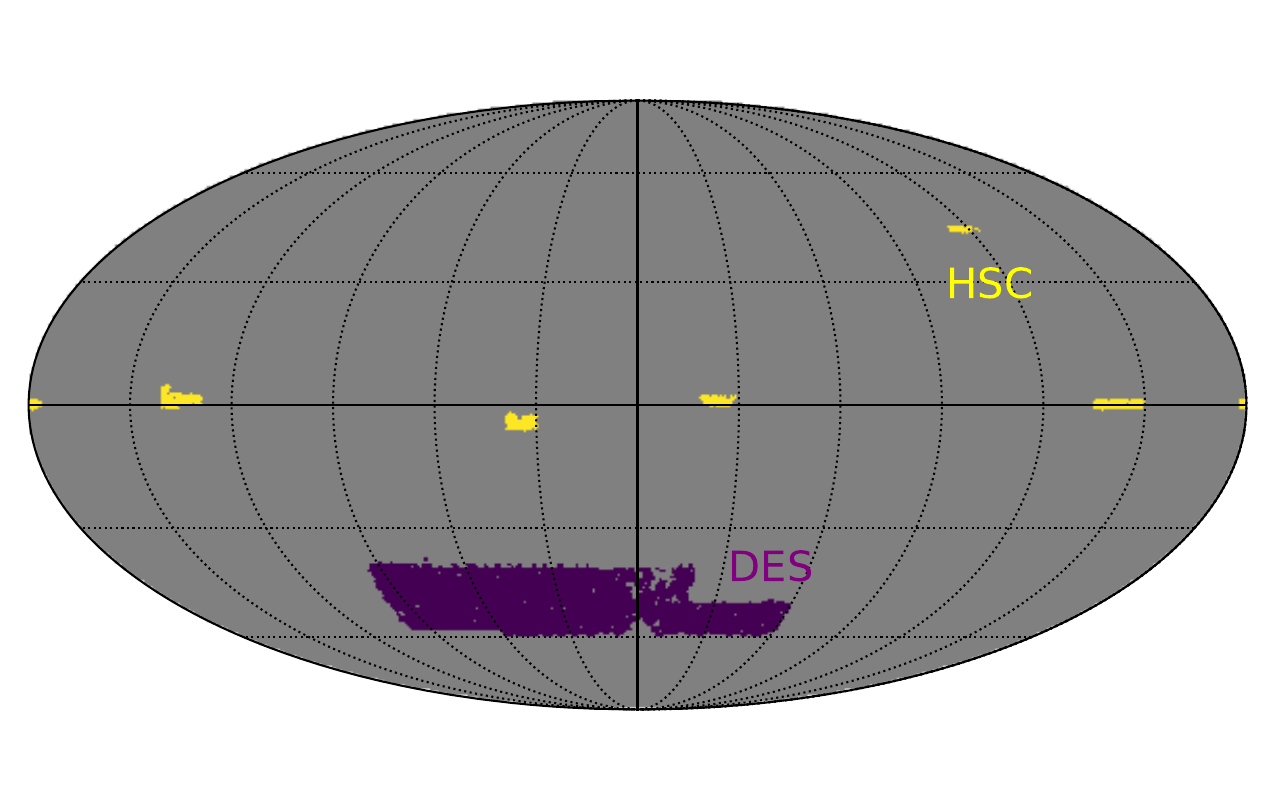}
      \caption{Sky footprint of the \desyo (purple) and \hscdro (yellow) samples. As they occupy disjoint sky regions, all statistical correlations between them come from the use of a common sample to obtain calibrated redshift distributions.}
      \label{fig:footprint}
    \end{figure}

    In this paper we combine the weak lensing datasets from the first year of observations of the Dark Energy Survey (DES) and Hyper Suprime-Cam surveys. We use the data produced in \cite{2010.09717} with two modifications: we calibrate the DES sample, instead of using the official $p(z)$, and we analytically marginalize over the $p(z)$ uncertainty at the covariance level as described in Section \ref{ssec:meth.nz_stat}. The spatial distribution of the two datasets are shown in Fig. \ref{fig:footprint}. As can be seen, the samples used for the cosmological analysis are completely disjoint on the sky, and are therefore uncorrelated except for their use of a common redshift calibration sample.

    The Hyper Suprime-Cam Subary Strategic Program (HSC SSP) is a wide-field imaging survey that will eventually cover $\unit[1000]{deg^2}$ in five filter bands ($grizy$). In this work we use the data from the first year of observation (HSC DR1) of the Wide layer, which covers $\sim \unit[108]{deg^2}$ and has a limiting magnitude of $\mlim = 26.4$ and a median $i$-band seeing of $\sim \unit[0.6]{arcsec}$. We choose cosmic shear sample, with a magnitude limit of $\mlim=24.5$, as defined in Ref.~\cite{1705.06745}. We use the photo-$z$ estimates using the {\tt Ephor\_AB} method to split the data into four \zbins with edges [0.3, 0.6, 0.9, 1.2, 1.5] (as done in the \hscdro analysis). We use the official galaxy distributions $p(z)$ used in \cite{1809.09148}. These were calibrated with \cosmos and a DIR method with weights given by self-organizing maps and can be found in Figure~\ref{fig:Nz}.

    The Dark Energy Survey is a 5-year survey that has observed $\unit[5000]{deg^2}$ in five different filter bands ($grizY$).  Observations were carried out from the Cerro Tololo Inter-American Observatory (CTIO) with the 4m Blanco Telescope, using the 570-Mpix Dark Energy Camera (DECam). The first data release used in this paper \citep{1708.01533} covers $\unit[1786]{deg^2}$ before masking. We use the \mcal weak lensing sample, divided in the four different redshift bins defined in \cite{1708.01532}. As in the case of \hscdro, \desyo also used COSMOS15 in the calibration process, for which they used a modified version of the BPZ method~\cite{astro-ph/9811189, 1708.01532}. In order to use a consistent methodology to obtain calibrated measurements of the redshift distribution across both \hsc and \des, we made alternative measurements of the \desyo redshift distributions via DIR using the \cosmos catalog, as described in Section \ref{ssec:results.pz}.

  \subsection{Power spectrum measurements}\label{ssec:data.cls}
    Our analysis is based on the shear power spectrum measurements of the \desyo and \hscdro datasets, presented and described in detail in \cite{2010.09717}. The power spectra were calculated via the the pseudo-$C_\ell$ method using \nmt \cite{1809.09603}. The \des power spectra were estimated from maps of the shear field constructed using the \healpix pixelisation scheme with resolution parameter $N_{\rm side}=4096$ (corresponding to a pixel size of about 1 arcmin). Given the smaller footprint size, the HSC power spectra were computed for each field independently using the flat-sky approximation, and co-added afterwards. The shape noise contribution to the auto-correlations was subtracted analytically as described in \cite{2010.09717}.
    
    The covariance matrix of these power spectra was estimated analytically, using the methods described in \cite{1906.11765,2010.09717}. The disconnected (Gaussian) component was calculated using the improved narrow-kernel approximation (NKA) \cite{2010.09717}, which is able to account for the impact of survey geometry on both the signal and noise components of the covariance. To the disconnected component we added the connected non-Gaussian part and super-sample contributions, estimated analytically using the halo model. Since they cover disjoint regions of the sky, we assumed no covariance between the HSC and DES power spectra. As we will see in Section \ref{sec:results}, the use of the same calibration sample (\cosmos) induces a covariance between both measurements after marginalising over the $p(z)$ uncertainties.
    
     When obtaining cosmological constraints, we use the angular power spectra up to $\ell = 2000$, and remove $\ell < 30$ for \desyo , and $\ell < 300$ for \hscdro. These scales were shown to be free of systematics in \cite{2010.09717} for \desyo and in the official analysis of \hscdro~\cite{1809.09148}. Note, however, that the official DES Y1 harmonic space analysis had much more conservative constraints based on the impact of baryonic effects in the angular power spectra~\cite{2111.07203}, which were not directly looked at in \cite{2010.09717}. 
     This high-$\ell$ cut corresponds to the scales used in the HSC-DR1 analysis, which were found to be largely unaffected by baryonic effects. Note, nevertheless, that this does not guarantee that baryonic effects are irrelevant for the joint analysis of both datasets (as hinted at in ~\cite{2111.07203}), and therefore a more thorough quantification of the impact of baryons would be necessary to ensure the robustness of the combined cosmological constraints. It is, however, reassuring that we are able to obtain a good $\chi^2$ for a model without baryonic effects. The power spectra and covariance matrices produced in \cite{2010.09717} and used in this analysis are publicly available\footnote{https://github.com/xC-ell/ShearCl}.

\section{Results}\label{sec:results}
  \subsection{$p(z)$ calibration from COSMOS}\label{ssec:results.pz}
    \begin{figure}
        \centering
        \includegraphics[width=0.8\textwidth]{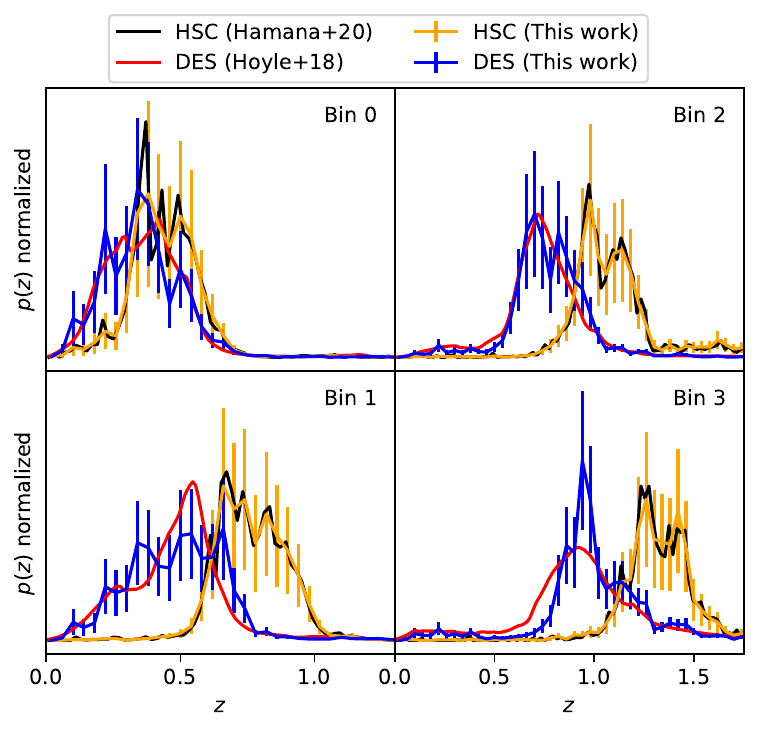}
        \caption{Redshift distributions for the DES (blue) and HSC (orange) samples calibrated with COSMOS and using DIR. In red, the official DES $p(z)$ and in black, those of HSC. Note that the difference between the two $N(z)$ for HSC is just the binning. The errors on the $p(z)$ have been estimated following Section~\ref{ssec:meth.nz_stat}. We find good agreement with the official results, except for the fourth bin where the shape changes significantly. As will be discussed in Section~\ref{ssec:results.single}, this leads to a small change in the inferred cosmological constraints. The mean redshifts can be found in Table~\ref{tab:z_mean}.}
        \label{fig:Nz}
    \end{figure}
    
    In order to ensure consistency in the analysis of the DES and HSC samples, we use the DIR method, as described in Section \ref{ssec:meth.nz_cal}, to estimate their redshift distributions and their associated uncertainties. As a calibrating sample ($S$ in the notation of Section \ref{ssec:meth.nz_cal}), we use the \cosmos catalog \cite{1604.02350}. This sample was also used in the analysis of \desyo, in combination with clustering redshifts and pdf stacking, and for \hscdro, using self-organising map to obtain colour-space weights. 

    In the case of \hscdro, a cross-matched catalog over the COSMOS field is provided with the public data release, including the lensing weights associated with each galaxy, as well as the SOM weights that correct for the different color distributions. We use this catalog directly when estimating the \hsc $p(z)$s.
    
    In the case of \des, we first cross-match the \cosmos catalog with the \des data over the same patch to associate each cross-matched source with fluxes measured in the \des bands. We used the four \textit{griz} bands and also checked that using the three \textit{riz} bands yielded negligible changes in the redshift distribution. We then apply the DIR procedure using the cross-matched catalog as $S$, and the main sample used for the cosmic shear analysis as $P$. It is worth noting that the colour distribution of the \desyo sample in the COSMOS field is noticeably different from that of the main sample, which results in a non-negligible shift in the inferred redshift distributions if we were to use it as $P$.
    
    \begin{table}[ht]
        \centering
        \begin{tabular}{|ccc|}
            \hline
             Bin &  DIR & Official\\
            \hline
             0 & $0.369 \pm 0.016$ & $0.389 \pm 0.016$\\
             1 & $0.503 \pm 0.017$ & $0.507 \pm 0.013$\\
             2 & $0.774 \pm 0.013$ & $0.753 \pm 0.011$\\
             3 & $1.004 \pm 0.015$ & $0.950 \pm 0.022$\\
            \hline
        \end{tabular}
        \caption{DES mean redshifts. We find that our calibration yields results compatible with the official ones.}
        \label{tab:z_mean}
    \end{table}
    
    The resulting redshift distributions are shown in Fig. \ref{fig:Nz} in blue (\des) and orange (\hsc), together with their statistical uncertainties, estimated as described in Section \ref{ssec:meth.nz_stat}. The figure also shows the official redshift distributions used in the \desyo (red)\footnote{Taken from \url{https://desdr-server.ncsa.illinois.edu/despublic/y1a1_files/}} and \hscdro (black)\footnote{Taken from \url{http://gfarm.ipmu.jp/~surhud/\#hamana}} analysis. The figure shows appreciable differences between the DIR estimate of the $p(z)$ and the official \desyo estimate, which combined pdf stacking and clustering redshifts.Looking at the mean redshifts in Table~\ref{tab:z_mean} we see deviations on the higher \zbins of order $\sim 2\sigma$ with shifts $\Delta z\simeq0.02$ and $\simeq0.05$, respectively. As we will see in Section \ref{ssec:results.single}, this results in a non-negligible shift in the posterior constraints on cosmological parameters. This effect was pointed out in \cite{2003.10454}, and will be discussed in further detail later. On the other hand, the differences in the \hsc $p(z)$ are just caused by a different binning choice and has no impact on our constraints (see Section \ref{ssec:results.single}).

    In order to compare our constraints with those found in the official \des and \hsc analyses, we need to ensure that the level of uncertainty in the redshift distributions is commensurate in both analyses. As has become standard in the analysis of cosmic shear data, the uncertainty on the $p(z)$ calibration was quantified in the official analyses of both datasets by introducing a free nuisance parameter for each redshift bin, characterising a shift $\Delta z$ in the mean of each redshift distribution, so that $p(z) \rightarrow p(z + \Delta z)$. This likely incorporates the main response of the weak lensing kernel to variations in the $p(z)$ \cite{1708.01532, 1809.09148}. To associate our estimate of the redshift distribution uncertainties with an effective prior uncertainty on $\Delta z$, we generate $10^4$ samples of the $p(z)$, sampling from a multivariate Gaussian distribution with mean $\bar{\bN}$ and covariance ${\sf P}$ (see Section \ref{ssec:meth.nz_stat}). For each of these, we compute the mean redshift $\enangle{z}$, and the prior on $\Delta z$ is then given by the standard deviation of $\enangle{z}$ over all samples. The mean and standard deviation of $\enangle{z}$ for each redshift bin in the DES sample are compared with the priors used in the official DES analysis in Table \ref{tab:priors}. In order to match the level of prior uncertainty used by DES, we adjusted our $p(z)$ uncertainties by a factor $\sqrt{2}$ (i.e. $f_P=2$ in Section \ref{ssec:meth.nz_stat}) when using the bias evolution from Eq.~\ref{eq:bg} and $\sqrt{3}$ when using the bias obtained from the cross-correlation with the CMB convergence field. We apply the same factor $\sqrt{3}$ to the HSC $p(z)$ uncertainties.

    \begin{figure}
        \centering
        \includegraphics[height=2.8in]{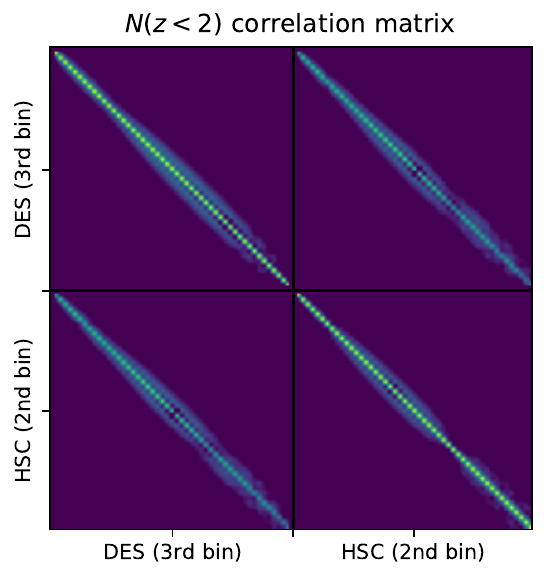}
        \includegraphics[height=2.8in]{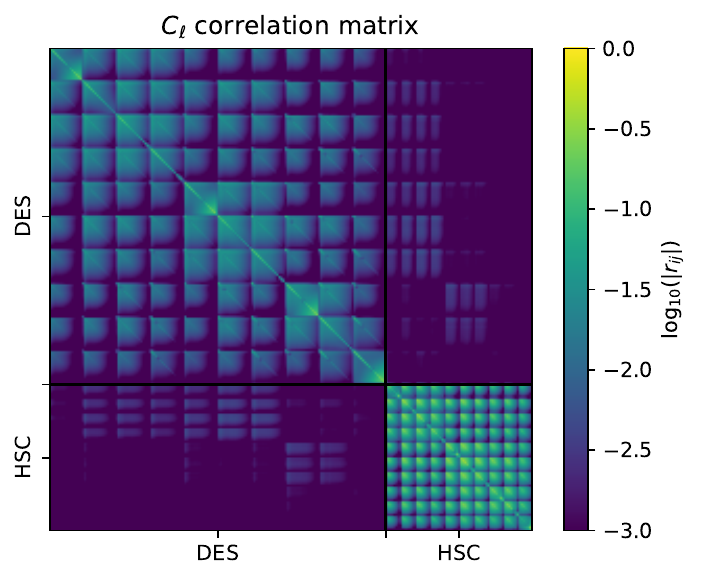}
        \caption{\ul{Left}. Redshift distribution correlation matrix for the third and second redshift bins of DES and HSC, respectively. The off-diagonal terms show that the calibration of the data sets of both surveys with the same sample (COSMOS) introduce correlations between the different $p(z)$. \ul{Right}. Covariance matrix of all angular power spectra including the contribution from the analytical marginalization of the $p(z)$. This introduces the visible correlations between HSC and DES data.}
        \label{fig:Nz_cov}
    \end{figure}

    The left panel of Fig.~\ref{fig:Nz_cov} shows the correlation matrix (see Eq.~\ref{eq:P}) for the redshift distributions of the third DES bin and the second HSC bin, which overlap significantly in redshift. There is a non-negligible correlation between the uncertainties of both redshift distributions, due to the use of a common calibrating sample. This then leads to a non-trivial correlation between the DES and HSC data vectors (see Eq.~\ref{eq:cov}) after marginalising over the $p(z)$ uncertainties. This is shown in the right panel of the same figure. The cross-terms between the DES and HSC blocks of the covariance are entirely due to the uncertainties in the redshift distribution estimation.
    
  \subsection{Characterising residual $p(z)$ systematics}\label{ssec:results.pz_syst}
    It has been argued that, in spite of their high accuracy, the photometric redshifts of the \cosmos catalog may induce a coherent shift in the redshift distribution of samples calibrated from it, particularly at high redshifts \cite{2007.11132}. This could cause a bias in the inferred cosmological parameters, and therefore this residual systematic must be modelled and marginalised over.
    
    \begin{figure}
        \centering
        \includegraphics[width=0.7\textwidth]{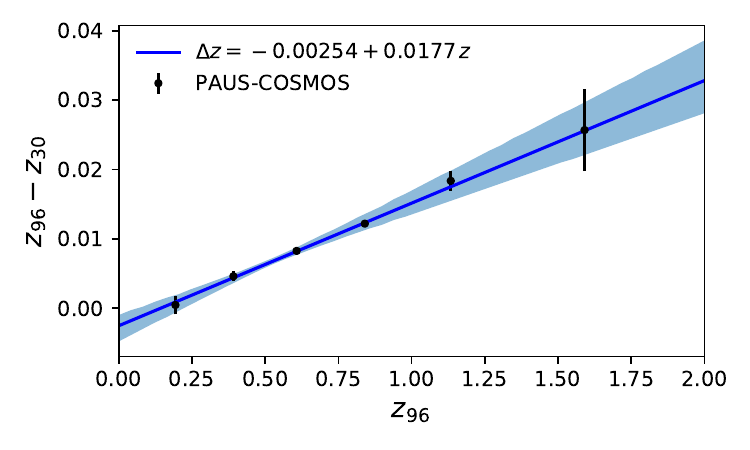}
        \caption{Redshift measurement difference between PAU and COSMOS15 with the best fit model (Eq.~\ref{eq:zdiff}) and its 68\% C.L. region. The points at $z\sim 0.5$ mainly drive the fit given the small error they have. These small errors are caused by a better agreement between PAU and \cosmos and a larger number of common galaxies lying at those redshifts.}
        \label{fig:PAUvsCOSMOS}
    \end{figure}
    
    To characterise this effect in the redshift distribution, we used the COSMOS catalog of \cite{2007.11132},  which combines the 30-band photometry of the original \cosmos catalog with 66 additional bands from the PAU survey \cite{1811.02368}. We retrieve the redshift obtained from the original \cosmos photometry ($z_{30}$) and the redshift using the combined photometry ($z_{96}$) for all cross-matched sources, and use them to measure the mean difference $\Delta z=z_{96}-z_{30}$ in 6 bins of redshift between $z=0$ and $z=1.6$. We estimate the uncertainty on the mean $\Delta z$ from the scatter of objects in each bin (normalised by $\sqrt{N}$, where $N$ is the number of objects). We then use these measurements to fit a model of the form
    \begin{equation}\label{eq:zdiff}
      \Delta z= A + B z_{96},
    \end{equation}
    to the measured $\Delta z$. The covariance of the two free parameters $(A,B)$ is then estimated using bootstrap resampling. The result is $A = -0.0025 \pm 0.0019$ and $B = 0.0177 \pm 0.0035$, with a correlation coefficient $r_{AB} = -0.96$\footnote{Note that we could have fitted $\Delta z = A + B (z_{96} - 0.5)$ to reduce the correlation between the coefficients to $r_{AB} \sim -0.3$ and the error of $A$. With this new fit $A = 0.00629 \pm 0.00047$, but the slope is left unchanged with $B = 0.0177 \pm 0.0038$. For simplicity we used the parametrization of Eq.~\ref{eq:zdiff}}. This result is illustrated in Fig. \ref{fig:PAUvsCOSMOS}, which shows the measured values of $\Delta z$ as a function of redshift, together with the best-fit linear model and the bootstrap-estimated errors. We notice that the fit is mainly driven by the points at $z\sim 0.5$ which have small error bars due to a combination of better agreement between PAU and \cosmos and a larger number of common galaxies in those bins. Furthermore, as noted in \cite{2007.11132}, the redshift of sources in COSMOS15 is biased low at high redshifts, and slightly high at low redshifts.

    To marginalise over this uncertainty, we proceed as in the standard cosmic shear analyses, shifting the redshift distributions of all samples by $\Delta z$, which now is a function of $z$. Note that the normalisation of the shifted probability distribution changes by a constant factor of $1-B$.

    We will explore the impact on our final results of various choices when marginalising over $(A,B)$. First, since the \desyo and \hscdro samples are not the same galaxy population, the systematic shift in their redshift distributions is not necessarily 100\% correlated. We will therefore produce constraints using the same $(A,B)$ parameters to describe the shift in DES and HSC (the limit of perfect correlation), and using two sets of parameters to describe each shift independently (the limit of no correlation). Secondly, we will study the effect on the final constraints of broadening the prior on $(A,B)$ obtained above.

  \subsection{Validation of the individual surveys}\label{ssec:results.single}
    \begin{figure}
        \centering
        \includegraphics[width=0.9\textwidth]{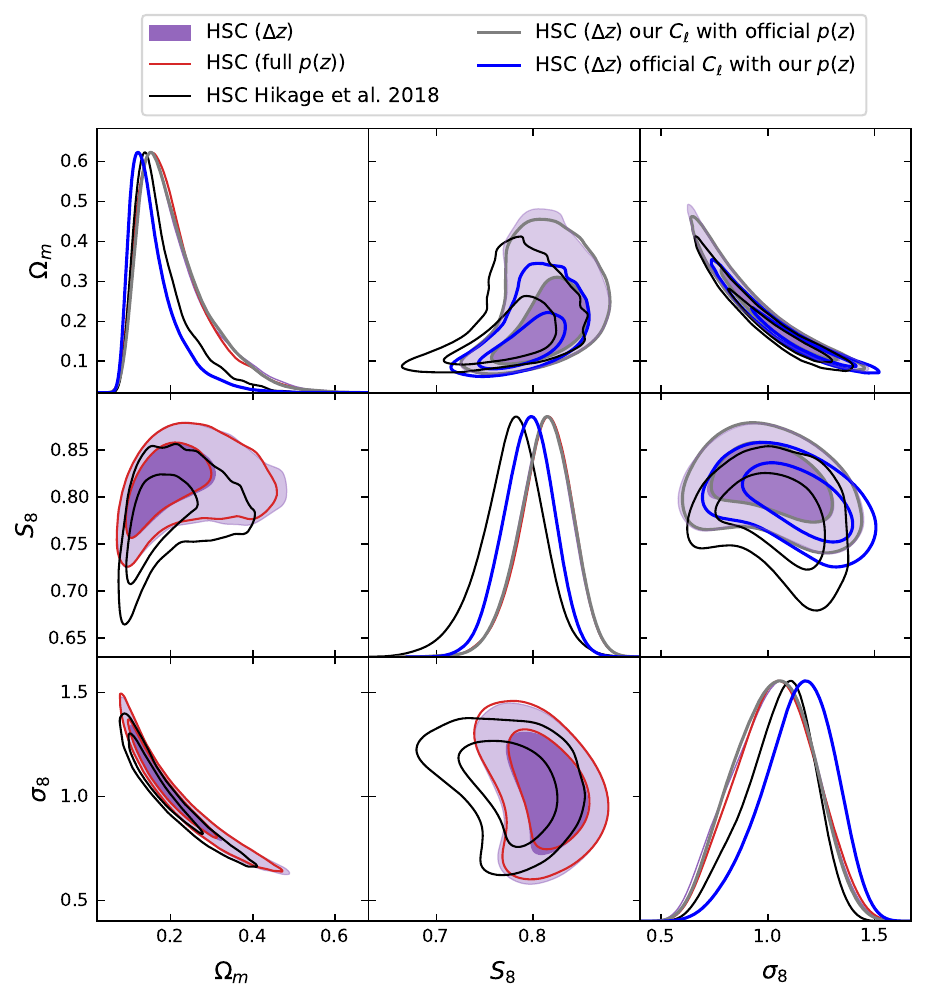}
        \caption{HSC posterior distributions. The \ul{lower triangle} shows the effect of using the priors in Table~\ref{tab:priors}. In purple, the result marginalizing at the likelihood level over redshift shifts $\Delta z$. In red, when the uncertainty on the $p(z)$ is analytically incorporated into the covariance. The agreement between both is very good. The reason why the contours with the shifts are slightly broader is probably due to a convergence issue, since they are not as smooth as in the analytically marginalized case. The \ul{upper triangle} focuses on the origin of the $S_8$ shift. In order to explore the origin of the shift we substitute, at a time, our $C_\ell$ or $p(z)$ by the official ones. The results are shown in blue (official $C_\ell$ but our $p(z)$) and gray (our $C_\ell$ but the official $p(z)$). Effectively, we find that changing the $p(z)$ does not affect the constraints, whereas using the official $C_\ell$ largely removes the shift. The remaining difference probably comes from the different modeling of the bandpower window functions applied to the theory $C_\ell$ in our likelihood, the different choice of sampler (Nested Sampling in their case and Metropolis-Hasting in ours), or the impact of the small PSF leakage terms in the theory prediction that we ignored here.}
        \label{fig:HSC_check}
    \end{figure}
    Before trying to obtain combined constraints from \desyo and \hscdro, we compare the results we obtain on each of these datasets individually with the official results published by both collaborations.

\begin{figure}
        \centering
        \includegraphics[width=\textwidth]{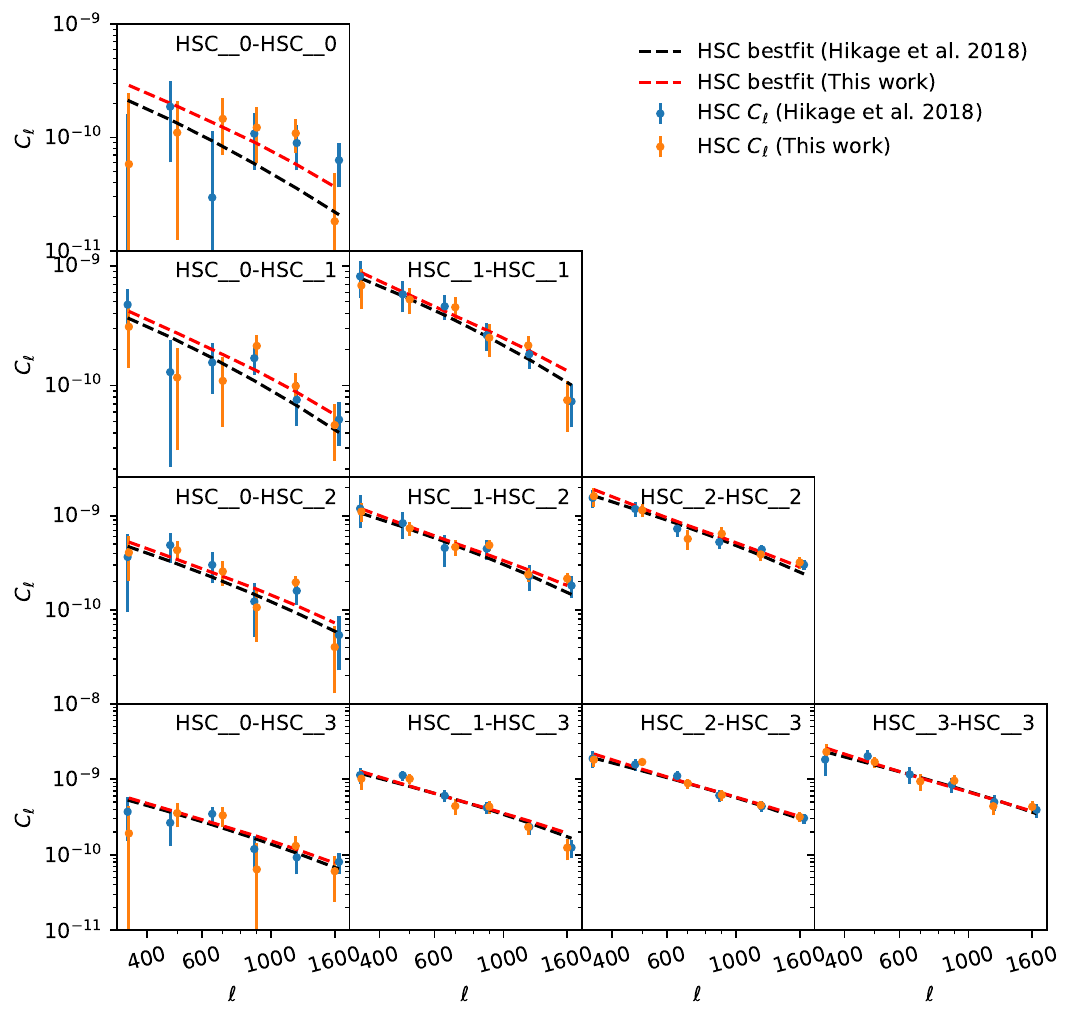}
        \caption{HSC data and bestfit angular power spectra. We see that our $C_\ell$  (orange) are compatible with the official ones (blue). Furthermore, the difference between bestfit power spectra obtained with the official data (black) and our data (red) appears mainly at the lower redshift bins, where the error bars are larger. In both cases we have a good fit to the data with $p=0.80$ with the official $C_\ell$ and $p=0.11$ with our $C_\ell$.}
        \label{fig:HSC_Cell}
    \end{figure}

    In the {\bf case of HSC}, most of our analysis choices are relatively similar to those made in the analysis of \cite{1809.09148}. We make use of angular power spectra calculated on the flat-sky, albeit binned into different bandpowers, calculated from maps generated with different pixelisation, and using a different (but mathematically equivalent) method to estimate the noise bias. We use the re-weighted COSMOS catalog made available by HSC to calibrate the redshift distributions of all four bins, although we sample them at a different number of redshifts (shown in Fig. \ref{fig:Nz}). We apply the same cosmological parameter priors used in \cite{1809.09148} with two exceptions: we allow for the multiplicative bias of each redshift bin to vary (as opposed to a single effective multiplicative parameter $m_{\rm eff}$ in \cite{1809.09148}), and we do not marginalise over parameters characterising the PSF leakage (although the effect of these was found to be small in \cite{1809.09148}). Fig. \ref{fig:HSC_check} shows the constraints on $\Omega_{\rm m}$, $\sigma_8$, and $S_8$ in five cases:
    \begin{itemize}
      \item The red contours show the constraints obtained from our data using the analytical marginalisation over the redshift distribution uncertainties described in the previous sections.
      \item In purple, we show the constraints obtained from our data marginalising over a shift in the mean of the redshift distribution of each redshift bin, using the priors prescribed in \cite{1809.09148}.
      \item The black contours show the official constraints of \cite{1809.09148}.
      \item In gray, the constraints when we use our $C_\ell$ but the official $p(z)$
      \item In blue, the constraints with the official $C_\ell$ but our $p(z)$
    \end{itemize}
    Our constraints are broadly compatible with the official ones, although we observe a small upwards shift on $S_8$, which aligns them more with the real-space results of \hscdro\cite{1906.06041, 2022PASJ...74..488H}. The size of this shift\footnote{We estimate the shifts in the paper by $\Delta\sigma_{12} = |x_1 - x_2| / \sigma_{x_1}$, where $x_1$ is the mean value of a given parameter (e.g. $S_8$) for the measurement 1 and $\sigma_{x_1}$ is its error estimated as the standard deviation of the parameter $x_1$ (we assume the posterior is Gaussian). For the case of estimating tensions where the measurements are uncorrelated, we use $\Delta\sigma_{12} = |x_1 - x_2| / \sqrt{\sigma_{x_1}^2 + \sigma_{x_2}^2}$.} is $\sim 1\sigma$ and comes from the differences in the $C_\ell$ binning described above. In order to show this, we run two different test cases: one with our $C_\ell$ and the official $p(z)$ and another with our $p(z)$ but the official $C_\ell$\footnote{Obtained from \url{gfarm.ipmu.jp/\~surhud/\#hikage}}. The results are shown in the upper triangle of Figure~\ref{fig:HSC_check}. We see that using the official $p(z)$ or ours does not change the posterior distributions, whereas the shift largely decreases when using the original $C_\ell$. The remaining difference probably comes from the different modeling choices at the level of the bandpower window functions that are applied to the theory $C_\ell$ in the likelihood, the use of different sampling methods (Nested Sampling~\cite{0704.3704, 0809.3437, 1306.2144} in their case and Metropolis-Hasting in ours), or the impact of the PSF leakage contribution to the theoretical predictions that we have neglected here. However, as shown in Figure~\ref{fig:HSC_Cell}, our $C_\ell$ are broadly in agreement with the official ones, and the differences between them do not display any specific trend. Furthermore, we see that the best fit $C_\ell$ do only significantly differ at lower redshift where the error bars are larger. Overall, we find that the bestfit $\chi^2 = 71.5$ ($p=0.11$) with our data vector, in contrast to $\chi^2 = 48.8$ ($p=0.80$) with the official $C_\ell$ and our $p(z)$.

    More interestingly, we find that the constraints found marginalising over the mean shifts $\Delta z$ are essentially equivalent to those found after fully marginalising over the $p(z)$ uncertainties using the method of \cite{2007.14989}. This reinforces the conclusion that the most relevant $p(z)$ uncertainty mode for cosmic shear is the redshift distribution mean, which affects the width and amplitude of the lensing kernel \cite{2003.11558,2206.10169}.

    In the {\bf case of DES}, our analysis choices differ more from those of \cite{1708.01538}. First, we use power spectra instead of real-space correlation functions. Secondly, we follow the cosmological parameter priors of the HSC analysis. The most significant differences in this case are that we use flat priors on $\log_{10}A_s$ instead of $A_s$, and that we sample over the physical dark-matter and baryon densities ($\Omega_{c/b}h^2$) instead of the total matter and baryon fractions ($\Omega_{m/b}$). Third, if we compare with \cite{2111.07203}, we also go to much smaller scales following \cite{2010.09717}, which we consider safe given the good $\chi^2$ obtained in that work and the lack of obvious systematics, such as $B$-modes. Finally, and most importantly, we use fiducial redshift distributions estimated from the \cosmos catalog using direct calibration, rather than the official redshift distributions made available by \cite{1708.01532}. The official redshift distributions were obtained by sampling the individual photo-$z$ probability densities of individual galaxies. Uncertainties on this fiducial distribution were then parametrised in terms of a shift parameter $\Delta z$, with a prior estimated from a combination of the reweighted \cosmos catalog and clustering redshifts. As discussed in detail in the re-analysis of the \desyo data carried out by \cite{1906.09262}, estimating the redshift distributions of the sample using direct calibration on various spectroscopic samples can lead to significant downward shifts in the posterior distribution of $S_8$. In \cite{1906.09262}, this was ascribed to potential systematics in the photometric redshifts of \cosmos, which could systematically underestimate the true shift of the fiducial \desyo distributions. Irrespective of this discussion, we chose to re-estimate the redshift distributions of the \desyo sample from \cosmos alone using DIR in order to be able to estimate consistently the statistical uncertainties of the distributions for both DES and HSC, including the correlations between both.

    \begin{figure}
        \centering
        \includegraphics[width=0.9\textwidth]{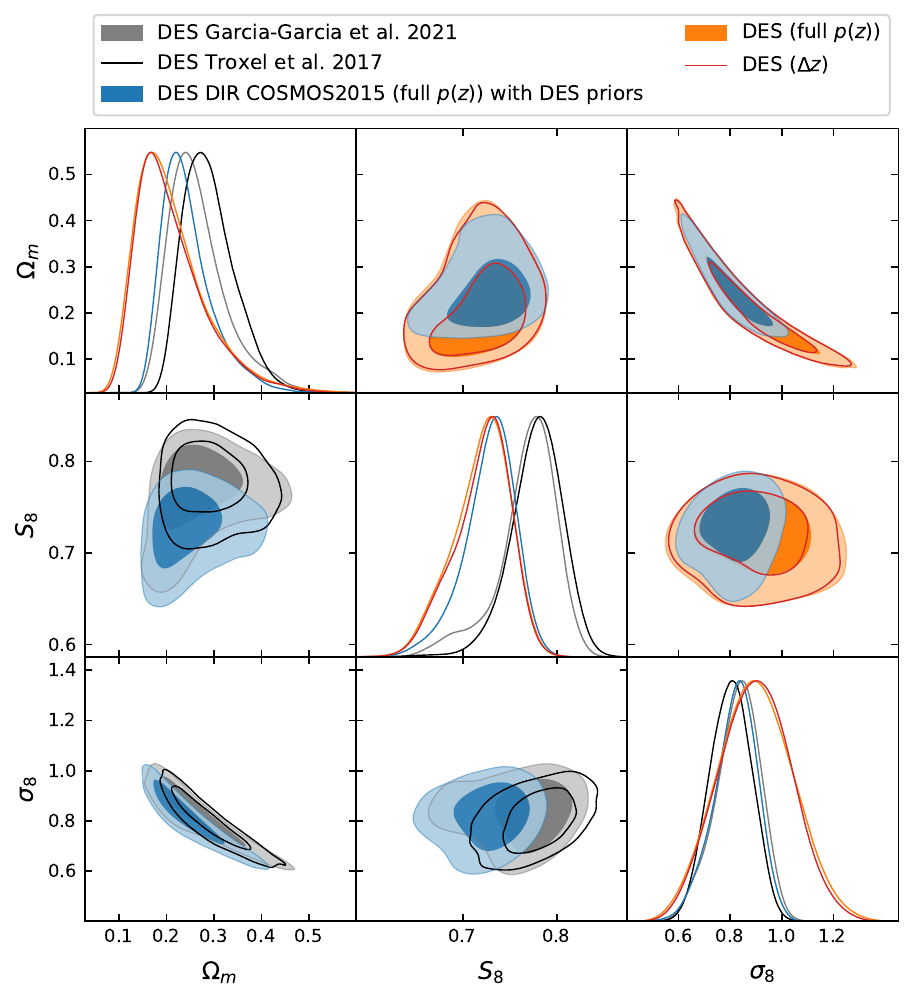}
        \caption{DESY1 posterior distributions for different analysis choices. The \ul{lower corner} shows the effect of the calibration of DES done in this work (blue). For comparison, we have plotted the official results (in black) and the contours obtained in using the data vector of this work but with the official $p(z)$ and marginalizing over a redshift shift at the likelihood level, as in Ref.~\cite{2105.12108} . Since the latter two agree, the shift must come from the calibration of this work. Note that, however, this shift is consistent with what was found using spectroscopic samples in \cite{1906.09262}. The \ul{upper corner} shows constraints using the HSC priors (see Table~\ref{tab:priors}). These are in orange and red. The orange contours are the fiducial result of this work and are from the case with the $p(z)$ uncertainty incorporated into the covariance matrix. In the case of the red contours, these are the result of marginalizing at the likelihood level over redshift shifts $\Delta z$. We see that these manage to encapsulate most of the $p(z)$ uncertainty but, still, the constraints are slightly tighter than those with the analytical marginalization.}
        \label{fig:DES_check}
    \end{figure}
    
    In order to distinguish between differences due to the prior choice and those due to the difference in the redshift distributions, we have first produced cosmological parameter constraints using the same priors as the official \desyo analysis. These are shown in the lower-left half of Fig. \ref{fig:DES_check}. The black contours show the official constraints\footnote{Taken from \url{https://desdr-server.ncsa.illinois.edu/despublic/y1a1_files/}}. The gray contours show the constraints found from the power spectrum measurements used here but employing the official \desyo redshift distributions and marginalising over the shift parameters as in \cite{2105.12108}. Finally, the blue contours show the constraints found from our power spectrum measurements and the redshift distributions found via DIR from \cosmos, marginalising over $\Delta z$, as in the two previous cases. When using the same redshift distributions, we find that our constraints are in good agreement with those found by \cite{1906.09262}. In turn, using the DIR $p(z)$s from \cosmos leads to a non-negligible downward shift of $\sim 1.3 \sigma$ in $S_8$. This is in agreement with the results of \cite{1906.09262}, and in fact we find that the DIR redshift distributions used here are in qualitative good agreement with those found by the authors of that paper using solely spectroscopic data\footnote{Private communication with Shahab Joudaki.}. It is worth noting that we carried out extensive tests to verify that our results were not sensitive to various choices made when estimating the DIR $p(z)$, including the number of nearest neighbors used in the spectroscopic catalog, the choice of photometric sample $P$ to derive weights (e.g. choosing the sample overlapping the \cosmos field as opposed to the cosmic shear sample), including the \mcal responsivity as a weight when estimating the redshift distribution, and varying the quality cuts made on \cosmos to define the $S$ sample.

    This leads us to two conclusions: firstly, the constraints from \desyo are highly sensitive to the methodology used to estimate the redshift distribution of the source sample. Secondly, the interpretation of the shift in $S_8$ found by \cite{1906.09262} in terms of photometric redshift systematics in \cosmos may be partially inaccurate, since the DIR $p(z)$s estimated from \cosmos are qualitatively similar to those of \cite{1906.09262}, and lead to a shift in $S_8$ similar to that found in the same paper. The origin of this disagreement then lies either in the method used to calibrate the official \desyo $p(z)$s, through stacking and clustering redshifts, or in the DIR approach used here to obtain calibrated $p(z)$s from the \cosmos data. As pointed out by \cite{2003.10454}, the fact that one has only four bands available might cause direct calibration methods to yield shifted distributions. Further work examining and quantifying the potential pitfalls of both methods will therefore be important for future analyses.

    The upper-right half of Fig. \ref{fig:DES_check} then explores the impact of the different choice of cosmological parameter priors used here. The red contours show the constraints found using the priors of Table \ref{tab:priors}, while still marginalising over $\Delta z$. Finally, the orange contours show the constraints found after analytically marginalising over the $p(z)$ uncertainties. As before, we find that the analytical marginalisation and the $\Delta z$ parametrisation lead to equivalent constraints. More importantly, the impact of the choice of priors is to marginally increase the left tail of the posterior $S_8$ distribution, and to shift the marginalised posterior distribution of $\Omega_m$ downwards.

    \begin{figure}
        \centering
        \includegraphics[width=0.7\textwidth]{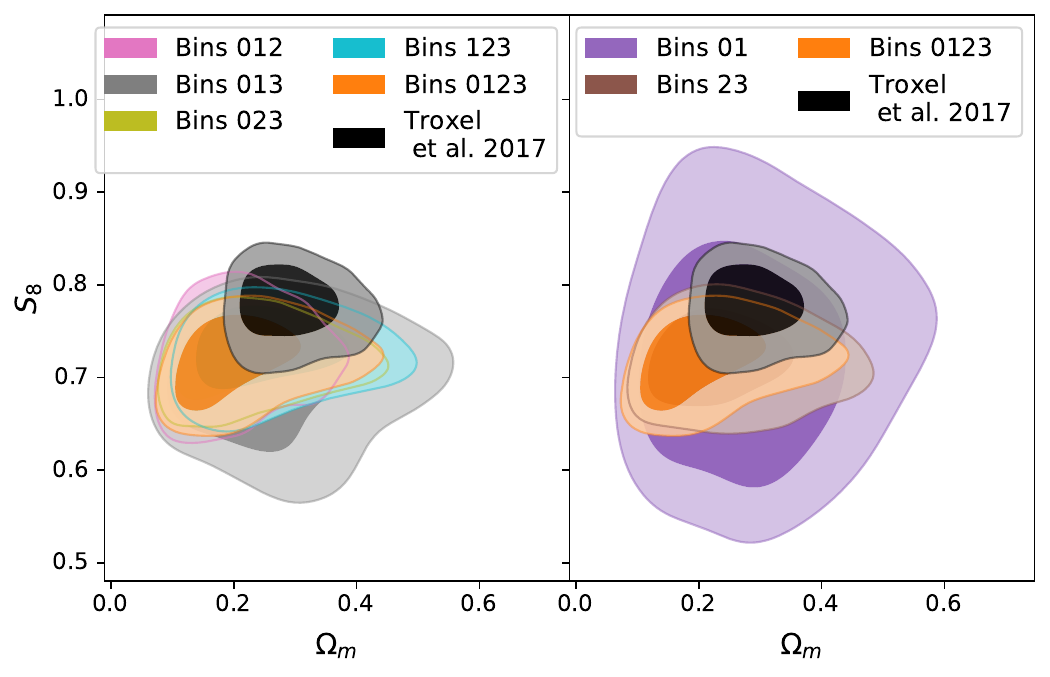}
        \caption{DESY1 posterior distributions considering only some of the redshift bins in the analysis. This exercise shows that the shift in $S_8$ is not coming from the calibration of the $p(z)$ of a particular tomographic bin, but all prefer lower values. This is consistent with what was found using spectroscopic samples in \cite{1906.09262}.}
        \label{fig:des_s8_shift}
    \end{figure}
    
    Before ending this section, we have carried out additional tests to try to ascertain the origin of the downwards shift in in $S_8$ in terms of the differences between the fiducial $p(z)$s and the DIR estimates used here, as shown in Fig. \ref{fig:Nz}. Fig. \ref{fig:des_s8_shift} shows, on the left-hand side, the constraints on $S_8$ and $\Omega_m$ found after removing each one of the four redshift bins (cyan, yellow, gray and pink contours in order of ascending redshift), together with the constraints found here (orange), and the fiducial \desyo constraints (black). We can see that the third redshift bin carries most of the constraining power, since the uncertainty on $S_8$ increases significantly when removing it. However, in all cases we find a posterior $S_8$ that is consistently lower than the official constraints. The right panel of the same figure shows the same constraints using only the first two (purple) or the last two (brown) redshift bins. The shift in $S_8$ is mostly caused by the two higher redshift bins, which also carry most of the constraining power. This result is in qualitative agreement with Fig. \ref{fig:Nz}, which shows that the DIR $p(z)$s in the two high redshift bins are more heavily weighted towards higher redshifts than the fiducial ones, and by table \ref{tab:z_mean}, which shows that the mean redshift of both beans is consistently higher (by $\Delta z\simeq0.02$ and $\simeq0.05$ respectively) than the official $p(z)$s.

  \subsection{Combining DES and HSC}\label{ssec:results.comb}
    \begin{figure}
        \centering
        \includegraphics[width=0.8\textwidth]{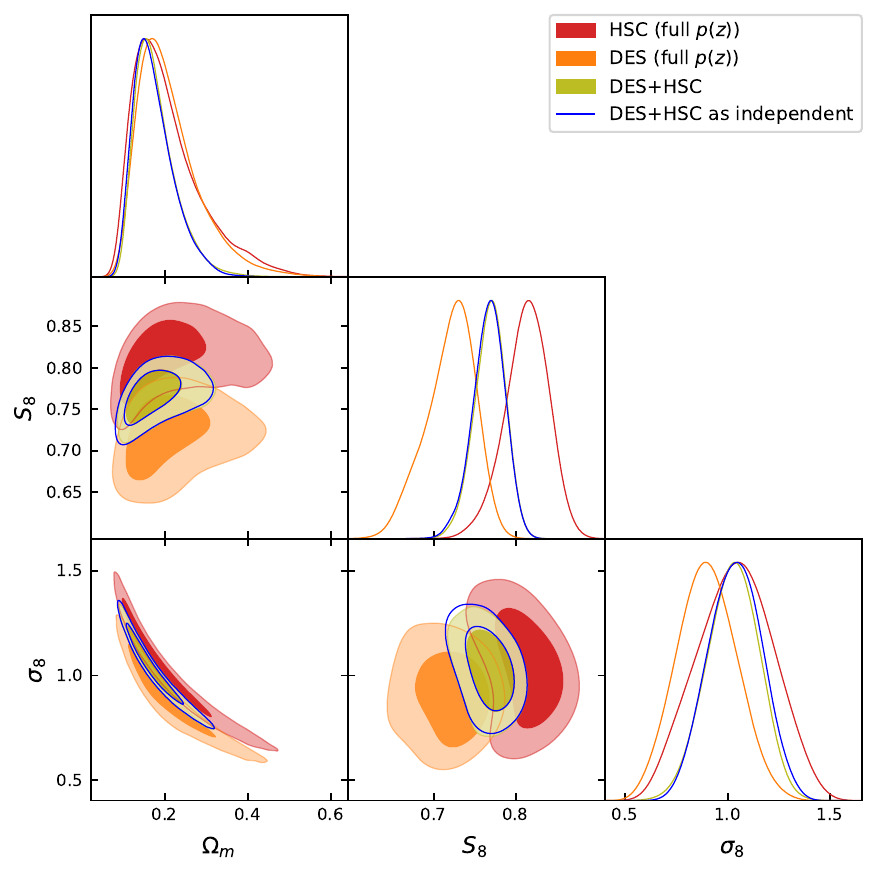}
        \caption{Posterior distributions for the DES (orange), HSC (purple) and their combination (pink and gray). The DES and HSC results differ by $\sim2.3 \sigma$, with DES prefering lower values of $S_8$ and $\sigma_8$. The gray contours show the combination of DES and HSC without taking into account the correlations from the redshift distributions calibration, whereas the pink contours do. Both constraints agree very well, showing that the correlation is a small, likely negligible effect.
       } 
        \label{fig:DES_HSC}
    \end{figure}
    We obtain combined constraints from \desyo and \hscdro by combining both data vectors together with their joint covariance matrix after analytically marginalising over the $p(z)$ uncertainties (see right panel of Fig. \ref{fig:Nz_cov}). The constraints resulting from this combination, together with the individual constraints of both surveys is shown in Fig. \ref{fig:DES_HSC}, and the constraints on $S_8$ from different analysis choices are listed in Table~\ref{tab:results}.

     \begin{table}
        \small
        \centering
        \def\arraystretch{1.2}
        \begin{tabular}{|lcccc|}
        \hline
                            &  $N_d$  & $\chi^2$ & $p$  & $S_8$ \\
        \hline 
        DES ($\Delta z$) & 240 & 234.48 & 0.55 & $ 0.720^{+0.036}_{-0.022}$ \\ 
        DES (full $p(z)$) & 240 & 235.15 & 0.54 & $ 0.720^{+0.036}_{-0.024}$ \\
        HSC ($\Delta z$) & 60 & 71.47 & 0.11 & $ 0.812^{+0.028}_{-0.025}$ \\
        HSC (full $p(z)$) & 60 & 71.80 & 0.11 & $ 0.813^{+0.030}_{-0.024}$ \\
        DES+HSC  & 300 & 311.70 & 0.28 & $ 0.768^{+0.021}_{-0.017}$ \\
        DES+HSC  as independent & 300 & 311.93 & 0.28 & $ 0.767^{+0.021}_{-0.018}$ \\
        DES+HSC  with systematic & 300 & 312.09 & 0.28 & $ 0.767^{+0.021}_{-0.018}$ \\
        DES+HSC  with independent systematic & 300 & 311.92 & 0.28 & $ 0.767^{+0.021}_{-0.018}$ \\
        DES+HSC  with systematic (prior x4) & 300 & 311.33 & 0.29 & $ 0.767^{+0.021}_{-0.017}$ \\
        Planck 2018 & -- & -- & -- & $0.832 \pm 0.013$ \\
        \hline
        \end{tabular}
        \caption{Main results. All results are shown with 68\% C.L. errors. The effective number of degrees of freedom is estimated as $N_{\rm dof,eff}=N_d-2$, where $N_d$ is the total number of data points.}
        \label{tab:results}
      \end{table}
    
    As noted in the previous section, the \desyo data favour a lower $S_8$ than \hscdro. Taking the central values and standard deviations of both datasets, and dividing by their errors added in quadrature, we find that the two estimates are in tension at the $\sim 2.3 \sigma$ level. This is traditionally not considered strong statistical evidence of tension for two virtually independent datasets, and therefore we will proceed to combine them. However, this tension is sufficiently significant to warrant a careful comparison between the results from both datasets in future data releases, in order to determine whether the difference is due to a statistical fluctuation, or evidence of potential systematics. We must note that, as shown in Table \ref{tab:results}, the joint best-fit model is able to fit the \desyo and \hscdro data simultaneously with a reasonable $\chi^2$ probability (28\%), which reinforces the conclusion that both datasets are not significantly incompatible.
    
    The combined constraint, shown as yellow contours in Fig. \ref{fig:DES_HSC}, lies in between both posteriors, and yields
    \begin{equation}\label{eq:final}
      S_8 = 0.768^{+0.021}_{-0.017}.
    \end{equation}
    The figure also shows the joint constraints found under the assumption that both datasets are completely independent (i.e. nulling the elements of the joint covariance matrix corresponding to power spectra from different surveys). We find that the constraints are almost completely unchanged ($S_8 = 0.767^{+0.021}_{-0.018}$). Thus, although present, the correlation between both datasets induced by their use of a common $p(z)$ calibration sample is completely negligible in the final cosmological constraints. As will be shown in Section~\ref{sec:forecast}, this will still be the case for Stage-IV surveys even though the final parameter uncertainties will greatly shrink.

\begin{figure}
        \centering
        \includegraphics[width=0.60\textwidth]{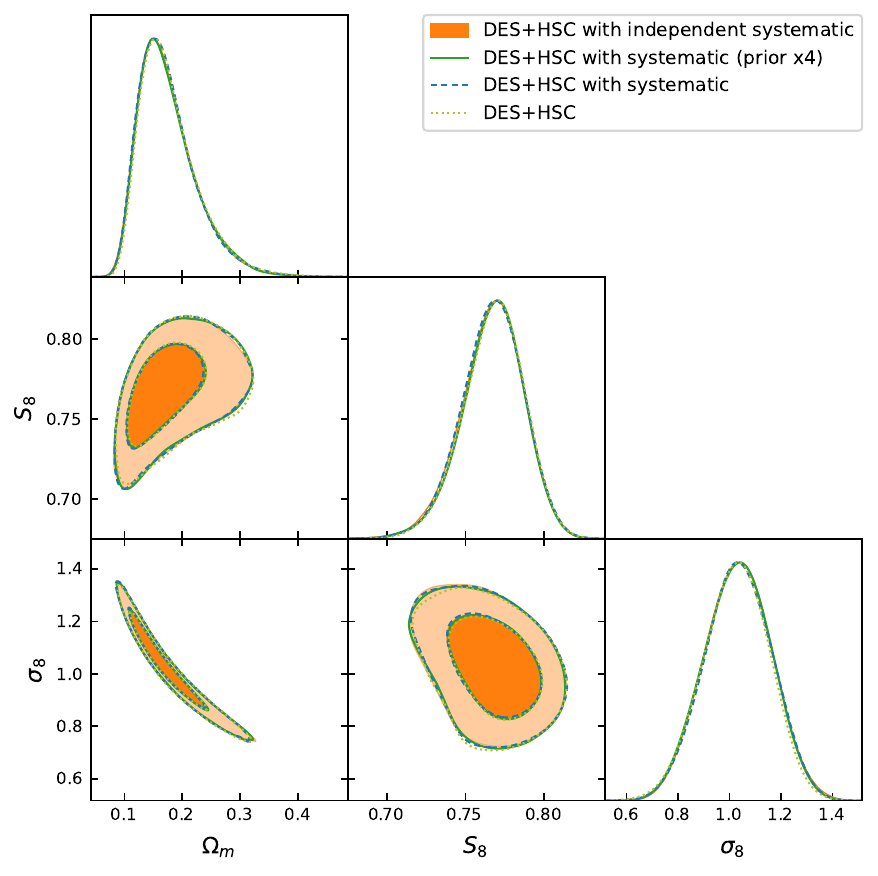}
        \includegraphics[width=0.39\textwidth]{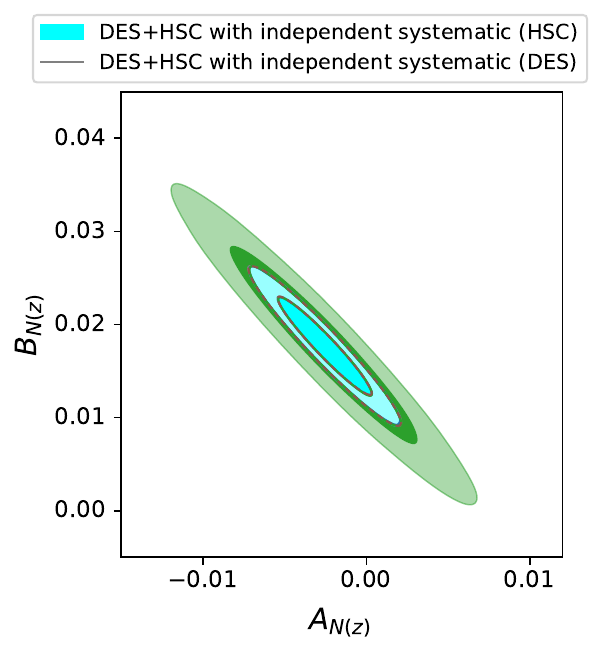}
        \caption{\ul{Left}. Posterior distributions for the combination of DES and HSC with (blue) and without (yellow) marginalizing over the calibration systematic bias. The effect is negligible with slightly broader contours in the cases with more parameters. This true even in the case we multiply the covariance matrix of the parameters systematic parameters by 4 (green) or when we use each survey has its own systematic bias (orange).  \ul{Right}. Posterior distributions for the calibration systematic bias parameters. The legend shows the color for the parameters constraints obtained in the case DES and HSC systematic parameters are different in the combined run (although it also includes the results from the cases on the left). The result is completely prior dominated in all cases, even for the case with the enlarged prior. The posteriors for all the different cases (except for the enlarged prior case) are the same and cannot be distinguished. This means that the calibration bias effect is small and that the cosmology information is not powerful enough to inform about possible redshift errors in the training sample.}
        \label{fig:DES_HSC_param}
    \end{figure}

    These results do not take into account the existence of a correlated systematic bias due to errors on the \cosmos photo-$z$. We account for these, as described in Section \ref{ssec:results.pz_syst}, by marginalising over $A$ and $B$ as defined in Eq. \ref{eq:zdiff} as additional nuisance parameters. The result is shown in Figure~\ref{fig:DES_HSC_param}. The yellow contours show the constraints in the absence of this additional systematic, while marginalising over $(A,B)$, assuming the same systematic shift in \desyo and \hscdro yields the blue contours. The constraints are almost equivalent, and lead only to a marginal ($<6\%$) increase in the uncertainties on $S_8$. Since the \desyo and \hscdro samples are not equivalent, the photo-$z$ systematic may not be the same in both, and therefore we consider the case where $A$ and $B$ are different parameters for each survey (i.e. we double the number of photo-$z$ nuisance parameters). The results are shown as orange contours, and are indistinguishable from the fully-correlated case. As in the case of the statistical uncertainties, we therefore conclude that the correlation between both samples due to the common use of \cosmos is negligible.
    
    In order to verify that this result is not a consequence of an under-estimate of the prior uncertainties on $A$ and $B$, we repeat the analysis doubling the width of the prior on these parameters (i.e. multiplying their covariance matrix by $4$). As in the previous case, this has no appreciable impact on the final constraints on $S_8$ (see green contours in Fig.~\ref{fig:DES_HSC_param}). As shown in  Fig. \ref{fig:PAUvsCOSMOS}, the size of this systematic is comparable to the shift in the mean of the redshift distributions allowed by the statistical uncertainties (see Table \ref{tab:z_mean}). Thus, the impact of this systematic may to some extent be already absorbed by the analytical marginalisation over the $p(z)$ statistical uncertainties.

\begin{figure}
        \centering
        \includegraphics[width=\textwidth]{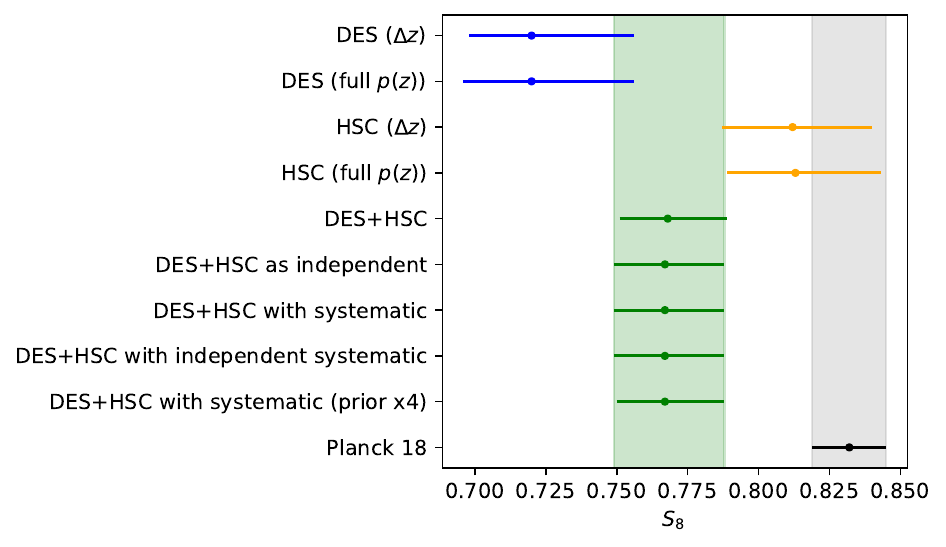}
        \caption{Cosmological parameter constraints for the studied cases here and \planck, for comparison. The error bars are the 68\%C.L. 1D-posterior intervals. We find that the lower values of $S_8$ respect to Planck, in this work, are mainly due to lower values of $\Om$. This is partially a volume effect, at least in the case of DES, since the sampling parameters in Table~\ref{tab:priors} allow much lower values of $\Om$ than the DES priors (see Figure~\ref{fig:DES_check}).}
        \label{fig:all_constraints}
    \end{figure}

    Our final combined constraint on $S_8$ is therefore given in Eq. \ref{eq:final}. This corresponds to a $\sim\sqrt{2}$ improvement in the statistical uncertainties compared to the errors of either survey. The constraints are in good agreement with those found by the latest cosmic shear analysis from the KiDS collaboration, and in tension with the value inferred by \planck at the $2.7\sigma$ level. This result, together with the constraints obtained from all other variations of our analysis, is also shown in Fig. \ref{fig:all_constraints}.

\section{Effect of correlated $p(z)$ uncertainties in future surveys}\label{sec:forecast}
  We have shown in the previous Section that the correlations induced by the redshift distribution calibration using a common spectroscopic sample have a negligible impact on the combined analysis of \desyo and \hscdro. This result, however, may no longer be valid with the advent of the Stage-IV surveys such as LSST~\citep{0805.2366,1809.01669} or Euclid~\citep{2012SPIE.8442E..0TL} which are expected to achieve percent-level accuracy on cosmological parameters and, thus, will become sensitive to modeling effects that can now be neglected.
  
  In this Section we explore the effect that the correlation between two future non-overlapping surveys will have in the parameter constraints. In order to model such a scenario, we produce a fiducial data vector given by the best-fit prediction found in the combined \desyo{}+\hscdro{} analysis, and artificially reduce the errors for these two surveys to match the expected observed area of $\sim\unit[10,000]{deg^2}$. This area is a factor of $\sim$10 and $\sim$100 larger than that covered by \desyo and \hscdro respectively. Taking into account that $\cov \propto \fsky^{-1}$, we modify the covariance matrices (before adding the analytically marginalized $p(z)$) so that
  \begin{equation}
      \cov^\text{\des} \rightarrow \cov^\text{\des} / 10 \qquad \cov^\text{\hsc} \rightarrow \cov^\text{\hsc} / 100 \,.
  \end{equation}
  To this modified covariance we then add the contribution from $p(z)$ uncertainties (Eq. \ref{eq:cov}), assuming the same prior uncertainty on the $p(z)$s of both experiments. This is likely a conservative assumption, since by the time these Stage-IV have finished taking their data, better calibrating spectroscopic samples will likely be available. Nevertheless, this allows us to quantify the potential impact of cross-survey correlations due to the $p(z)$ calibration in a conservative scenario.
  
    \begin{figure}
        \centering
        \includegraphics[width=0.70\textwidth]{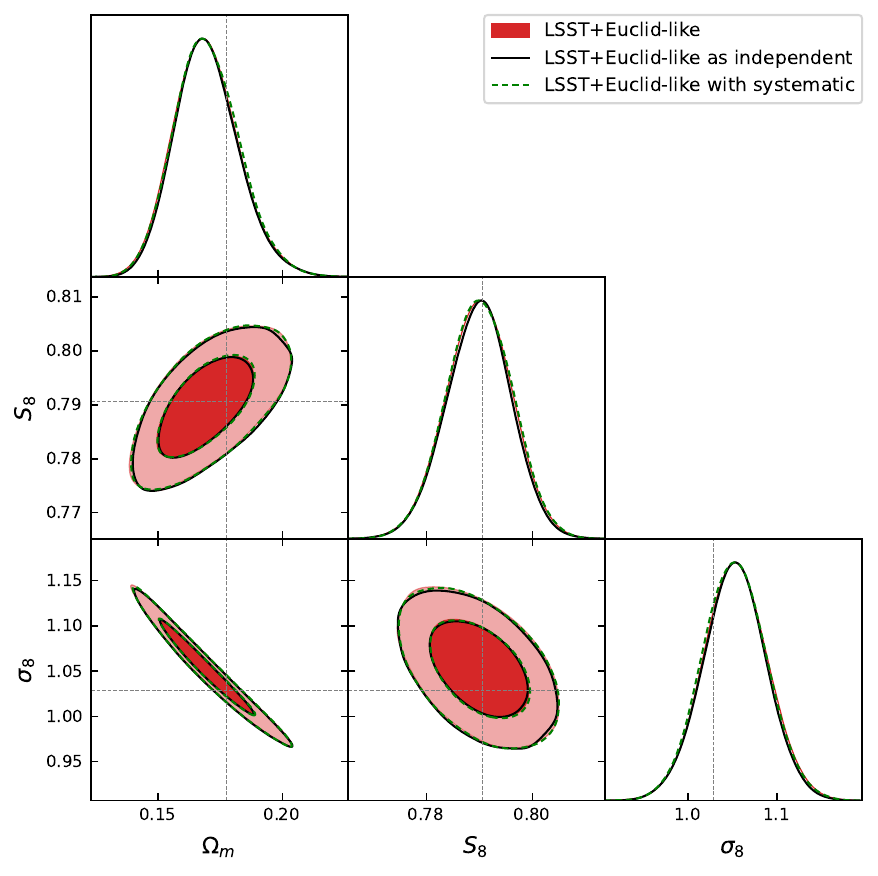}
        \caption{Posterior distributions for the combination of two non-overlapping LSST/Euclid-like surveys taking into account (red) or not (black) the correlations induced by the $p(z)$ calibration and including the systematic effect (Eq.~\ref{eq:zdiff}, green dashed line). We see that the effect of both the $p(z)$-induced correlations and the systematic is negligible with only slightly broader contours when the correlations are taken into account.}
        \label{fig:forecast}
    \end{figure}
  
  In order to test for the effect of the induced correlations we constrain the parameter space in two different scenarios: one with the covariance as explained above and another where we set the correlations between the two surveys induced by the $p(z)$ calibration to zero, while leaving the $p(z)$ contribution to the individual covariances intact. As can be seen in Fig.~\ref{fig:forecast}, both cases produce largely equivalent constraints 
  We find that this is also the case after accounting for the impact of correlated $p(z)$ systematics (Eq.~\ref{eq:zdiff}), which barely change the constraints. Therefore, we conclude that even in the case of Stage-IV surveys the effects of the correlated statistical uncertainties induced by the $p(z)$ calibration, and the systematic bias in the training sample will have a negligible effect. It is worth noting that this result is true for the redshift bins explored in the \desyo and \hscdro analyses, and other galaxy samples may display different levels of correlation between bins. The strong correlations present in the \desyo-\hscdro case (see Fig. \ref{fig:Nz_cov}) suggest this will likely not change the results found here.

\section{Conclusions}\label{sec:conc}
  In this paper we have addressed the problem of combining independent weak lensing datasets in the presence of correlated systematic uncertainties from the use of common external datasets for the purpose of redshift distribution calibration. To do so, we have proposed a method that combines the analytical marginalisation over statistical uncertainties in the reconstruction of the $p(z)$ from the common sample, followed by a marginalisation, at the likelihood level, over residual systematics in it. We have then applied this methodology to the \desyo and \hscdro samples with redshift distributions estimated from the \cosmos catalog, obtaining cosmological constraints from their joint analysis.

  The methodology used to characterise the correlated statistical uncertainties in the recovered redshift distribution, and to analytically marginalise over them, is based on the work of \cite{2007.14989}, proposed for the analysis of projected galaxy clustering, and which we have applied to cosmic shear data for the first time. In applying this methodology to HSC and DES, we have first recalculated the fiducial redshift distributions for the \desyo sample using direct reweighting (DIR) of the \cosmos catalog. This was done in order to obtain consistent $p(z)$ measurements and uncertainties for both datasets. The DIR redshift distributions thus obtained have noticeable differences with respect to the fiducial distributions used by \desyo, particularly in the two highest redshift bins, which dominate the cosmic shear signal. As noted in previous work \cite{1906.09262}, this leads to a sizeable shift in the posterior distribution of the ``clumping'' parameter $S_8$. Although this was originally ascribed to the use of \cosmos to calibrate the allowable shift in the mean of the fiducial redshift distributions used in the \desyo analysis, our result seems to hint towards a different motivation. Ultimately, the cause these differences may lie in the methods used to calibrate the $p(z)$s (pdf stacking, clustering redshifts, or the use of direct calibration with information from only 4 bands \cite{2003.10454}), rather than the dataset they were applied to. This might support the need for new methods, such as the shear-ratios, that are able to self-calibrate the weak lensing $p(z)$ uncertainties \cite{2105.13542}. The constraints from \desyo found here thus differ from the official ones by a $1.3\sigma$ downward shift in $S_8$.

  The individual constraints from \hscdro and \desyo are found to be in mild $\sim2.3 \sigma$ tension, although a joint best-fit model can be found that is a good fit to both datasets.  After combining both datasets, the uncertainties on $S_8$ improve by a factor of $\sim\sqrt{2}$, and the posterior constraints (Eq. \ref{eq:final}) are in tension with the \planck value for this parameter at the $2.7\sigma$ level. It is worth noticing, however, that although the scales used in this work have been shown to be largely unaffected by baryonic effects~\cite{1809.09148}, this might not be true for the combined analysis as hinted in~\cite{2111.07203} and in line with a possible reason of the $S_8$ tension as pointed out in \cite{2206.11794}. In any case, we have shown that the impact of the correlation between both datasets induced by the use of a common redshift calibration sample (\cosmos), is completely negligible in the current analysis, as is the additional marginalisation over a systematic shift in the mean redshifts of the \cosmos sample calibrated from the COSMOS-PAU catalog, regardless of whether the systematic is taken to be fully correlated or fully uncorrelated in \hscdro and \desyo. Additionally, we have shown that this will also be the case for next-generation Stage-IV datasets\footnote{Note that, in general, the overlapping area between the photometric surveys and the high-quality phonometric or spectroscopic surveys will be disconnected. In this case, the calibration will be highly uncorrelated with only remaining correlations coming from correlated systematics in the training sample such as the one investigated in Sections~\ref{ssec:meth.nz_syst} and~\ref{ssec:results.pz_syst}.}. This will largely simplify their analysis when different experiments make use of the same external catalogs for the purposes of redshift calibration.  

\newpage
\section*{Acknowledgements}
  We would like to thank Jo Dunkley, Shahab Joudaki, Emily Phillips Longley, and David N. Spergel for useful comments and discussions.
 
  CGG and PGF acknowledge support from European Research Council Grant No:  693024 and the Beecroft Trust. We made extensive use of computational resources at the University of Oxford Department of Physics, funded by the John Fell Oxford University Press Research Fund. DA is supported by the Science and Technology Facilities Council through an Ernest Rutherford Fellowship, grant reference ST/P004474. AN is supported through NSF grants AST-1814971 and AST- 2108126. CS is supported by a Junior Leader fellowship from the ”la Caixa” Foundation (ID 100010434), with code LCF/BQ/PI22/11910018, and by grants AST-1615555 from the U.S. National Science Foundation, and DE-SC0007901 from the U.S. Department of Energy (DOE). For the purpose of Open Access, the author has applied a CC BY public copyright licence to any Author Accepted Manuscript version arising from this submission.
  
  \ul{Software}:  We made extensive use of the {\tt numpy} \citep{oliphant2006guide, van2011numpy}, {\tt scipy} \citep{2020SciPy-NMeth}, {\tt astropy} \citep{1307.6212, 1801.02634}, {\tt healpy} \citep{Zonca2019}, {\tt matplotlib} \citep{Hunter:2007} and {\tt GetDist} \citep{Lewis:2019xzd} python packages. The codes used to produce these results can be found in \url{https://gitlab.com/carlosggarcia/DESxHSC}.
   
  This paper makes use of software developed for the Large Synoptic Survey Telescope. We thank the LSST Project for making their code available as free software at \url{http://dm.lsst.org}.
    
  \ul{Data:} 
  This project used public archival data from the Dark Energy Survey (DES). Funding for the DES Projects has been provided by the U.S. Department of Energy, the U.S. National Science Foundation, the Ministry of Science and Education of Spain, the Science and Technology Facilities Council of the United Kingdom, the Higher Education Funding Council for England, the National Center for Supercomputing Applications at the University of Illinois at Urbana-Champaign, the Kavli Institute of Cosmological Physics at the University of Chicago, the Center for Cosmology and Astro-Particle Physics at the Ohio State University, the Mitchell Institute for Fundamental Physics and Astronomy at Texas A\&M University, Financiadora de Estudos e Projetos, Funda{\c c}{\~a}o Carlos Chagas Filho de Amparo {\`a} Pesquisa do Estado do Rio de Janeiro, Conselho Nacional de Desenvolvimento Cient{\'i}fico e Tecnol{\'o}gico and the Minist{\'e}rio da Ci{\^e}ncia, Tecnologia e Inova{\c c}{\~a}o, the Deutsche Forschungsgemeinschaft, and the Collaborating Institutions in the Dark Energy Survey.
        
  The Collaborating Institutions are Argonne National Laboratory, the University of California at Santa Cruz, the University of Cambridge, Centro de Investigaciones Energ{\'e}ticas, Medioambientales y Tecnol{\'o}gicas-Madrid, the University of Chicago, University College London, the DES-Brazil Consortium, the University of Edinburgh, the Eidgen{\"o}ssische Technische Hochschule (ETH) Z{\"u}rich,  Fermi National Accelerator Laboratory, the University of Illinois at Urbana-Champaign, the Institut de Ci{\`e}ncies de l'Espai (IEEC/CSIC), the Institut de F{\'i}sica d'Altes Energies, Lawrence Berkeley National Laboratory, the Ludwig-Maximilians Universit{\"a}t M{\"u}nchen and the associated Excellence Cluster Universe, the University of Michigan, the National Optical Astronomy Observatory, the University of Nottingham, The Ohio State University, the OzDES Membership Consortium, the University of Pennsylvania, the University of Portsmouth, SLAC National Accelerator Laboratory, Stanford University, the University of Sussex, and Texas A\&M University.
    
  Based in part on observations at Cerro Tololo Inter-American Observatory, National Optical Astronomy Observatory, which is operated by the Association of Universities for Research in Astronomy (AURA) under a cooperative agreement with the National Science Foundation.
    
  The Hyper Suprime-Cam (HSC) collaboration includes the astronomical communities of Japan and Taiwan, and Princeton University. The HSC instrumentation and software were developed by the National Astronomical Observatory of Japan (NAOJ), the Kavli Institute for the Physics and Mathematics of the Universe (Kavli IPMU), the University of Tokyo, the High Energy Accelerator Research Organization (KEK), the Academia Sinica Institute for Astronomy and Astrophysics in Taiwan (ASIAA), and Princeton University. Funding was contributed by the FIRST program from Japanese Cabinet Office, the Ministry of Education, Culture, Sports, Science and Technology (MEXT), the Japan Society for the Promotion of Science (JSPS), Japan Science and Technology Agency (JST), the Toray Science Foundation, NAOJ, Kavli IPMU, KEK, ASIAA, and Princeton University. 
    
  The Pan-STARRS1 Surveys (PS1) have been made possible through contributions of the Institute for Astronomy, the University of Hawaii, the Pan-STARRS Project Office, the Max-Planck Society and its participating institutes, the Max Planck Institute for Astronomy, Heidelberg and the Max Planck Institute for Extraterrestrial Physics, Garching, The Johns Hopkins University, Durham University, the University of Edinburgh, Queen’s University Belfast, the Harvard-Smithsonian Center for Astrophysics, the Las Cumbres Observatory Global Telescope Network Incorporated, the National Central University of Taiwan, the Space Telescope Science Institute, the National Aeronautics and Space Administration under Grant No. NNX08AR22G issued through the Planetary Science Division of the NASA Science Mission Directorate, the National Science Foundation under Grant No. AST-1238877, the University of Maryland, and Eotvos Lorand University (ELTE) and the Los Alamos National Laboratory.
    
  Based in part on data collected at the Subaru Telescope and retrieved from the HSC data archive system, which is operated by Subaru Telescope and Astronomy Data Center at National Astronomical Observatory of Japan.
    
 \bibliography{main,non_ads}
 
\end{document}